\documentclass[11pt,oneside,a4paper,onecolumn]{elsarticle}

\usepackage{times}
\usepackage{fullpage}

\usepackage{gensymb}
\usepackage{alltt}
\usepackage{upquote}
\usepackage{float}
\usepackage{array}
\usepackage{pgfplots}
\usetikzlibrary{positioning}
\usepackage{graphicx}
\usepackage[toc,page]{appendix}
\usepackage[utf8]{inputenc}
\usepackage{amsmath,mathtools}
\usepackage{amsthm}
\usepackage{amsfonts}
\usepackage{amssymb}
\usepackage{enumitem}
\usepackage{caption}
\usepackage{subcaption}
\usepackage{xcolor}
\usepackage{scalerel}
\usepackage{algorithm}
\usepackage{algorithmic}

\usepackage{multicol}
\usepackage{multirow}
\usepackage{url}
\usepackage{booktabs}
\usepackage{nicefrac}

\usepackage{nowidow}

\usepackage{soul}
\usepackage{slashed}
\usetikzlibrary{patterns}
\usetikzlibrary{shapes.arrows}
\usetikzlibrary{fillbetween,backgrounds}
\usepackage{placeins}
\pgfplotsset{compat=newest}
\usepackage{diagbox}
\usetikzlibrary{arrows.meta}

\pgfdeclarelayer{nodelayer}
\pgfdeclarelayer{edgelayer}

\pgfsetlayers{background,nodelayer,edgelayer,main}

\usepackage[final]{hyperref} 
\hypersetup{
	colorlinks=true,       
	linkcolor=blue,        
	citecolor=blue,        
	filecolor=magenta,     
	urlcolor=blue         
}
\usepackage[capitalize]{cleveref}
\creflabelformat{equation}{#2\textup{#1}#3}

\newcolumntype{M}[1]{>{\centering\arraybackslash}m{#1}}
\newcolumntype{N}{@{}m{0pt}@{}}

\tikzset{>=latex}
\def\@author#1{\g@addto@macro\elsauthors{\normalsize%
    \def\baselinestretch{1}%
    \upshape\authorsep#1\unskip\textsuperscript{%
      \ifx\@fnmark\@empty\else\unskip\sep\@fnmark\let\sep=,\fi
      \ifx\@corref\@empty\else\unskip\sep\@corref\let\sep=,\fi
      }%
    \def\authorsep{\unskip,\space}%
    \global\let\@fnmark\@empty
    \global\let\@corref\@empty
    \global\let\sep\@empty}%
    \@eadauthor={#1}
}

\graphicspath{{Images/}}

\begin{document}
\begin{frontmatter}
\title{Predicting Flow-Induced Vibration in Isolated and Tandem Cylinders Using Hypergraph Neural Networks}
\author[ubc]{Shayan Heydari\corref{cor1}}
\ead{shayan.heydari@ubc.ca}

\author[ubc]{Rui Gao}
\ead{garrygao@mail.ubc.ca}

\author[ubc]{Rajeev K. Jaiman}
\ead{rjaiman@mech.ubc.ca}
\cortext[cor1]{Corresponding author}
\address[ubc]{Department of Mechanical Engineering, The University of British Columbia, Vancouver, BC V6T 1Z4}

\begin{abstract}
We present a finite element-inspired hypergraph neural network framework for predicting flow-induced vibrations in freely oscillating cylinders. The surrogate architecture transforms unstructured computational meshes into node-element hypergraphs that encode higher-order spatial relationships through element-based connectivity, preserving the geometric and topological structure of the underlying finite-element discretization. The temporal evolution of the fluid–structure interaction is modeled via a modular partitioned architecture: a complex-valued, proper orthogonal decomposition-based sub-network predicts mesh deformation using a low-rank representation of Arbitrary Lagrangian-Eulerian (ALE) grid displacements, while a hypergraph-based message-passing network predicts the unsteady flow field using geometry-aware node, element, and hybrid edge features. High-fidelity ALE-based simulations provide training and evaluation data across a range of Reynolds numbers and reduced velocities for isolated and tandem cylinder configurations. The framework demonstrates stable roll-outs and accurately captures the nonlinear variation of oscillation amplitudes with respect to reduced velocity, a key challenge in surrogate modeling of flow-induced vibrations. In the tandem configuration, the model successfully resolves complex wake-body interactions and multi-scale coupling effects, enabling accurate prediction of pressure and velocity fields under strong wake interference conditions. Our results show high fidelity in reproducing force statistics, dominant frequencies, and flow-field dynamics, supporting the framework’s potential as a robust surrogate model for digital twin applications.
\smallskip
\smallskip

 \end{abstract}
\end{frontmatter}


\section{Introduction}
\label{sec:intro}
\begin{figure}[t]
\centering
\includegraphics[width=0.6\linewidth]{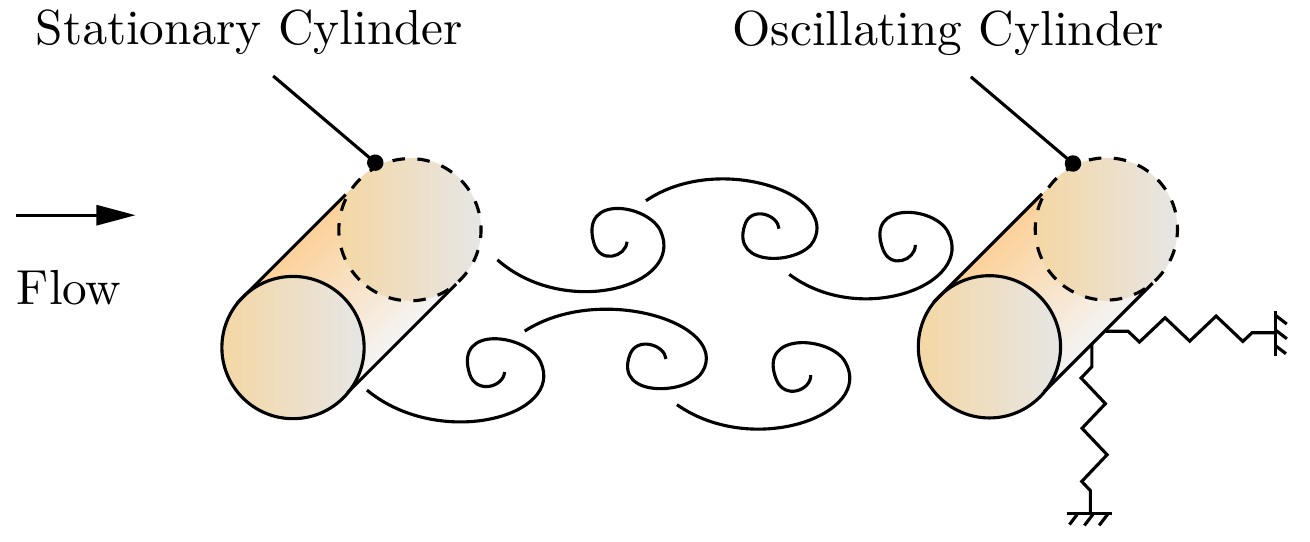}
\caption{Schematic of a freely oscillating cylinder placed in the wake of a rigid stationary cylinder, serving as a canonical problem for studying vortex-structure interactions.}
\label{motivation}
\end{figure}

Flow-induced vibration (FIV) of bluff bodies is a foundational problem in fluid-structure interaction (FSI) research~\cite{Jaiman_FIV, Paidoussis2013}, with critical implications for the safety and performance of numerous engineering systems, including offshore platforms~\cite{NARENDRAN2018111} and marine risers~\cite{joshi2017variationally}. Among these, the flow-induced response of cylindrical structures stands out due to its complex nonlinear behavior and broad practical relevance. A particularly well-studied case is the vortex-induced vibration of a circular cylinder subjected to cross flow. In such cases, the cylinder can undergo frequency lock-in~\cite{Williamson2004, Sarpkaya2004}, where the shedding frequency of alternating von Kármán vortices synchronizes with a natural frequency of the structure, resulting in sustained, large-amplitude transverse oscillations~\cite{blevins, heydari2022fluid}. In multi-body configurations, such as tandem cylinders, wake-induced vibration introduces additional complexity. Coherent vortices shed from the upstream body interact with the downstream structure (as shown in Fig.~\ref{motivation}) and give rise to intricate multi-frequency responses~\cite{ASSI_BEARMAN_MENEGHINI_2010, Heydari_Jaiman_PRF_2025}. Accurate modeling of these coupled dynamics remains vital for the design and operation of vortex-dominated systems.

Despite substantial advances in analytical~\cite{Mottaghi_Gabbai_Benaroya_2020} and numerical~\cite{jaiman2022computational} modeling, predicting FIVs remains a challenging task. The governing partial differential equations are nonlinear and strongly coupled, demanding resolution of multiple spatio-temporal scales in fluid and structural domains~\cite{en15228719}. While high-fidelity simulations using partitioned or monolithic schemes offer detailed insights into vortex-structure interactions~\cite{jaiman2022computational}, their computational cost often precludes real-time applications such as active control, system monitoring, or digital twin implementation.
To reduce computational demands, reduced-order modeling approaches have been developed to construct low-dimensional representations of complex dynamics. Techniques such as proper orthogonal decomposition~\cite{rowley2004model, miyanawala2019decomposition}, dynamic mode decomposition~\cite{SCHMID_2010}, the eigensystem realization algorithm~\cite{chizfahmERA}, and Galerkin projection~\cite{galerkin-noack-1994, NoackReview2011} have been applied to model key features of unsteady fluid flows and FSI systems. A comprehensive review of the relevant methodologies can be found in~\cite{rowley-rom}. 

Conventional reduced-order models have proven useful in many settings; however, their reliance on simplifying assumptions often renders them inadequate for capturing the strongly nonlinear and multi-scale phenomena characteristic of wake-body interactions. To address these limitations, machine learning approaches have emerged as promising alternatives for surrogate modeling of fluid mechanics systems~\cite{brunton2020machine, gupta2022hybrid, Herrmann_Kollmannsberger_2024}. Among recent advances in machine learning-based surrogate modeling, deep neural networks have shown strong potential for capturing the complex, high-dimensional spatio-temporal patterns that arise in fluid flows and fluid–structure interaction problems. These models can learn directly from the data, offering greater flexibility in representing nonlinear dynamics compared to conventional approaches. In general, deep learning methods can be categorized into two classes: purely data-driven models that infer mappings solely from data, and physics-informed neural networks that embed governing physical laws, such as the Navier-Stokes equations, into the learning process~\cite{RAISSI2019686}. Hybrid approaches that combine reduced-order models with deep learning have also been explored~\cite{bukka2021assessment, labzour2023pinnrom}. 

Among existing deep learning models, increasing interest has been directed toward spatially expressive architectures, most notably convolutional neural networks and graph neural networks (GNNs). These architectures are designed to more effectively learn spatial correlations, which are critical for resolving multi-scale phenomena and interactions in complex physical systems. Convolutional neural networks are shown to be effective in learning local spatial correlations on regular grids, making them well-suited for structured mesh data. For example, \cite{bukka2021assessment} developed a convolution-based recurrent autoencoder architecture that learns directly from structured flow field data to predict unsteady flow dynamics around two side-by-side cylinders. Although convolutional neural networks have proven to be effective in capturing spatial correlations, their use is most natural in domains with regular grid structures. Graph neural networks offer a flexible alternative by operating directly on unstructured meshes, representing the domain as a graph with nodes corresponding to mesh points and edges capturing topological connectivity~\cite{Ogoke_Meidani_Hashemi_Farimani_2021, GNNBook2022}. Using message passing and neighborhood aggregation, GNNs can efficiently learn from irregular spatial data, allowing them to be applied effectively in the surrogate modeling of geometrically complex and locally refined computational settings~\cite{gnn-review}.

\begin{figure*}[t]
\centering
\begin{subfigure}[b]{0.28\textwidth}\centering
\includegraphics[width=.85\linewidth]{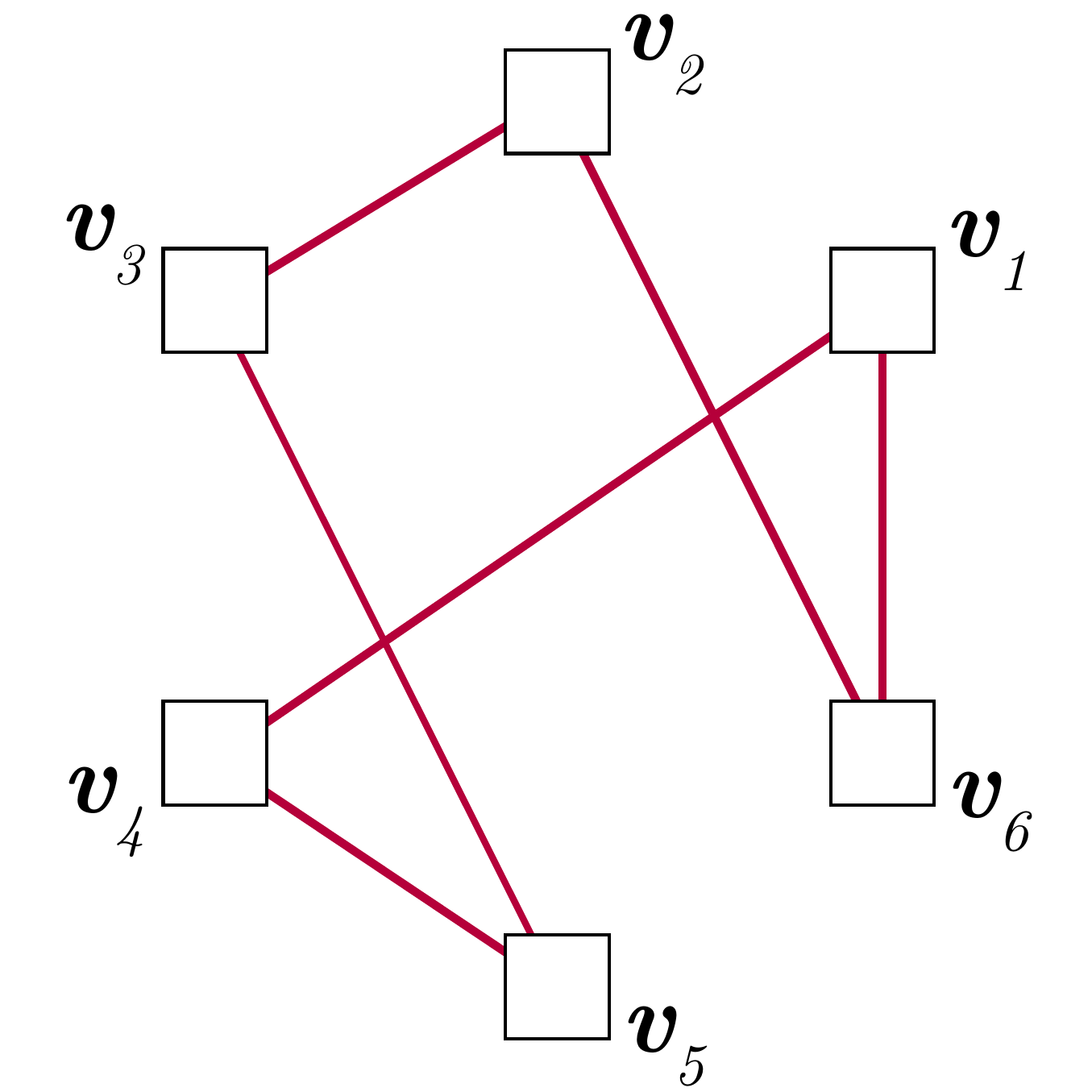}
\caption{}
\end{subfigure}
\begin{subfigure}[b]{0.28\textwidth}\centering
\includegraphics[width=.85\linewidth]{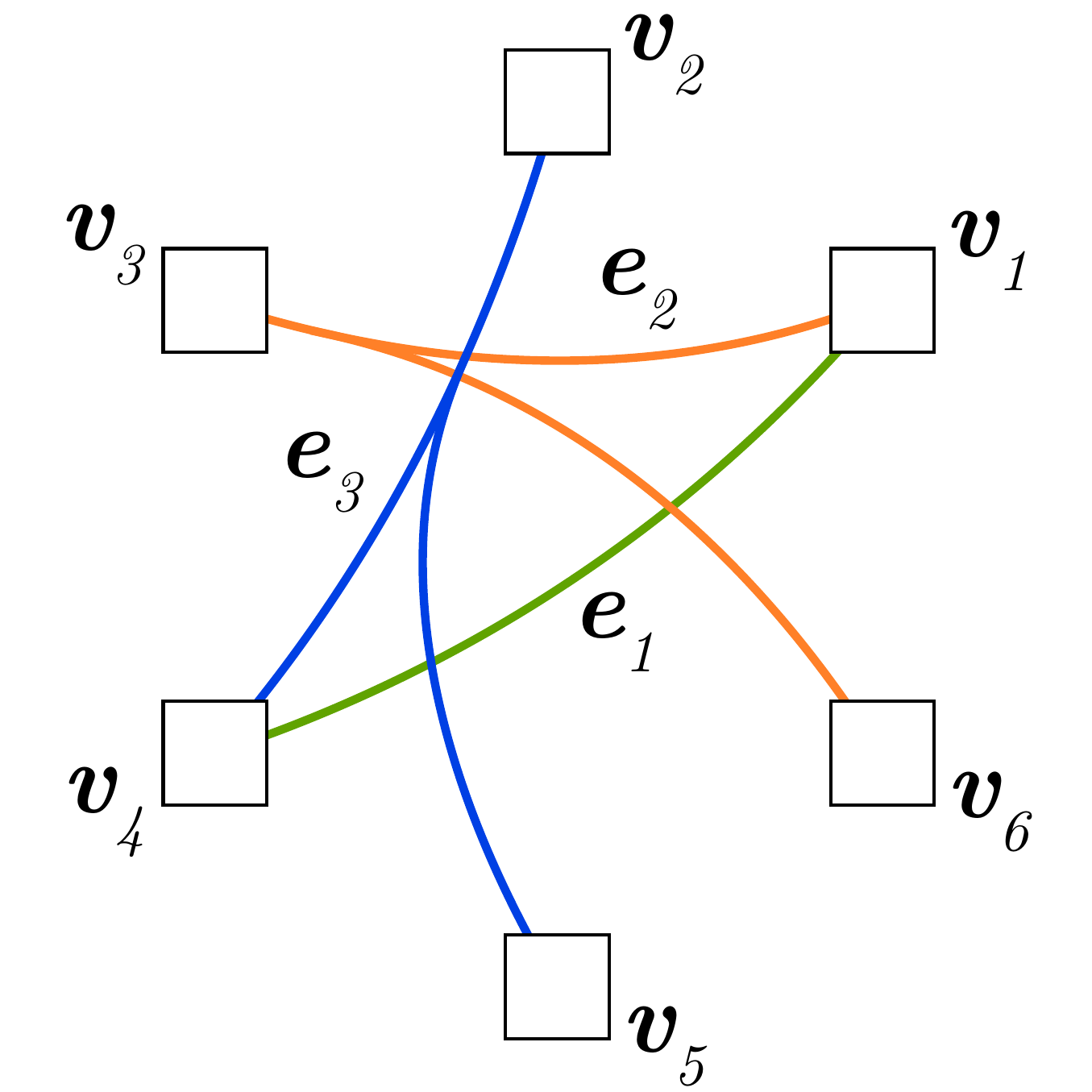}
\caption{}
\end{subfigure}
\begin{subfigure}[b]{0.28\textwidth}\centering
\includegraphics[width=.85\linewidth]{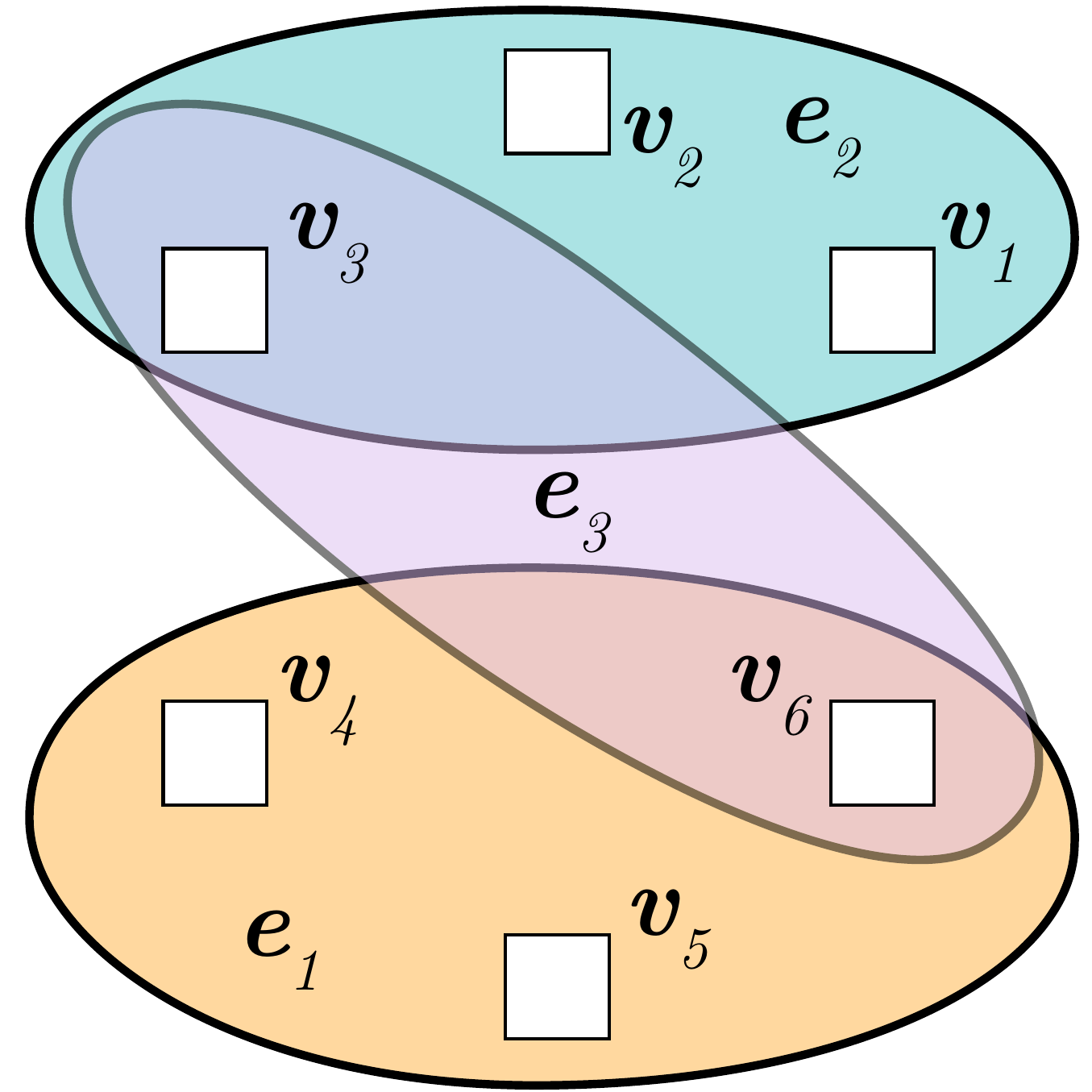}
\caption{}
\end{subfigure}
\caption{Comparison between a standard graph and a hypergraph. In the standard graph (a), each edge connects exactly two vertices ($v$). In contrast, (b) and (c) illustrate two typical representations of a hypergraph, where each hyperedge ($e$) can simultaneously connect multiple vertices.}
\label{comparison}
\end{figure*}

Hypergraph neural networks extend the expressive capacity of GNNs by generalizing edges to hyperedges, which can connect multiple nodes simultaneously within a single relation, as illustrated in Fig.~\ref{comparison}. This higher-order structure enables the encoding of complex multi-node interactions that GNNs cannot directly represent~\cite{dai2023hypergraph}. 
Recent studies have demonstrated the effectiveness of hypergraph neural networks in modeling mesh-based fluid dynamics~\cite{gao2024finite, GAO2024104858} and fluid–structure interaction systems~\cite{gao2024predicting}. These models exhibit strong flexibility, are well-suited to unstructured meshes, and offer the expressive capacity required to capture complex relational dependencies inherent in such systems. Extending efforts in hypergraph-based surrogate modeling, we present a hypergraph neural network framework to model flow-induced vibrations. The application of such surrogate models to vortex-induced and wake-induced vibrations remains limited and largely unexplored in a systematic manner. This work seeks to address that gap by evaluating the effectiveness of a hypergraph-based surrogate model in learning and predicting the spatio-temporal evolution of unsteady flow fields and structural responses in canonical FIV systems. By embedding higher-order interactions through node-element hypergraphs and decoupling the evolution of mesh and flow states, the framework is applied to benchmark configurations involving an elastically mounted rigid cylinder in isolated and tandem arrangements. 

The scenarios studied in this work are chosen so that we can assess the predictive performance of the framework under complex behaviors such as vortex synchronization and wake interference effects. The results demonstrate that the surrogate model achieves accurate rollout predictions across a range of non-dimensional parameters, including Reynolds number ($Re$) and reduced velocity ($U^*$). This work contributes to the development of mesh-consistent surrogates for fluid–structure interaction systems, with potential applications in digital twins and real-time monitoring in engineering contexts. The remainder of this paper is structured as follows. Section~\ref{sec:methodology} introduces the governing equations for the full-order modeling of fluid–structure interactions and details the hypergraph neural network framework. Section~\ref{sec:results} presents the numerical results along with a discussion of the performance of the model. Finally, Section~\ref{sec:conclusion} summarizes the main findings and provides directions for future work.

\section{Methodology}
\label{sec:methodology}
This section outlines the governing equations for fluid–structure interaction simulations, the discretization approach using the arbitrary Lagrangian–Eulerian formulation, and the construction of the node-element hypergraph. The architecture of the hypergraph neural network framework is also described, comprising two decoupled sub-networks: one for mesh deformation using a complex-valued proper orthogonal decomposition representation, and another for flow field prediction via message passing on the hypergraph.

\subsection{Full-order Model}
The viscous fluid flow is governed by the incompressible Navier–Stokes equations, formulated in the arbitrary Lagrangian–Eulerian reference frame over the time-dependent fluid domain \( \Omega^{f}(t) \), and expressed as follows:
\begin{equation}\label{DNS}
\rho^f \left( \frac{\partial \textit{\textbf{u}}^f}{\partial t} \bigg|_{\hat{x}} + \big(\textit{\textbf{u}}^f - \textit{\textbf{w}}\big) \cdot \nabla \textit{\textbf{u}}^f \right) = \nabla \cdot {\boldsymbol{\sigma}}^f + \textit{\textbf{b}}^f \quad \text{on } \quad \Omega^f(t),
\end{equation}
\begin{equation}\label{DNS-2}
\nabla \cdot \textit{\textbf{u}}^f = 0 \quad \text{on } \quad \Omega^f(t),
\end{equation}
where \( \textit{\textbf{u}}^f = \textit{\textbf{u}}^f(\textit{\textbf{x}}, t) \) and \( \textit{\textbf{w}} = \textit{\textbf{w}}(\textbf{\textit{x}}, t) \) denote the fluid and mesh velocities at each spatial point \( \boldsymbol{x}^{f} \in \Omega^{f} \), respectively. The term \( \textit{\textbf{b}}^f \) represents the body force per unit mass, and \( \boldsymbol{\sigma}^f \) is the Cauchy stress tensor for a Newtonian fluid, defined as:
\begin{equation}
\boldsymbol{\sigma}^f = -p\boldsymbol{I} + \mu^f \left( \nabla \boldsymbol{u}^f + (\nabla \boldsymbol{u}^f)^\mathrm{T} \right),
\end{equation}
where \( p \) denotes the fluid pressure and \( \boldsymbol{I} \) is the identity tensor. The first term in Eq.~\ref{DNS} corresponds to the partial time derivative of \( \boldsymbol{u}^{f} \) while the ALE referential coordinate, $\hat{x}$, is kept fixed.

A rigid structure mounted on an elastic support and immersed in fluid flow is subjected to hydrodynamic forces that can induce structural vibrations. The resulting translational motion of the structure about its center of mass along the Cartesian axes is described by the following equation:
\begin{equation}
\boldsymbol{m} \cdot \frac{\partial \boldsymbol{u}^s}{\partial t} + \boldsymbol{c} \cdot \boldsymbol{u}^s + \boldsymbol{k} \cdot \big( \boldsymbol{\phi}^s(\boldsymbol{z}_0, t) - \boldsymbol{z}_0 \big) = \boldsymbol{F}^s + \boldsymbol{b}^s \quad \text{on } \quad \Omega^s,
\end{equation}
where \( \boldsymbol{m} \), \( \boldsymbol{c} \), and \( \boldsymbol{k} \) denote the structural mass, damping, and stiffness, respectively. The fluid traction vector acting on the structure is given by \( \boldsymbol{F}^s \), and \( \boldsymbol{b}^s \) represents the body force. The structural domain is denoted by \( \Omega^s \), and \( \boldsymbol{u}^s(t) \) corresponds to the structural velocity. 

The fluid–structure coupling is enforced through the continuity of velocity and traction at the interface between the fluid and the structure. The motion of the rigid body is described by the position vector \( \boldsymbol{\phi}^s \), which maps the initial configuration \( \boldsymbol{z}_0 \) to its current position at time \( t \). Let \( \gamma \) be a Lagrangian point on the fluid–structure interface \( \Gamma \), and \( \boldsymbol{\phi}(\gamma, t) \) represent its mapped position as the rigid body moves. Since the motion of the rigid body and the surrounding flow field evolve continuously, the no-slip condition and traction continuity must be satisfied at the interface such that:
\begin{equation}
\boldsymbol{u}^f\big(\boldsymbol{\phi}^s(\boldsymbol{z}_0, t), t\big) = \boldsymbol{u}^s(\boldsymbol{z}_0, t),
\end{equation}
\begin{equation}\label{INTERFACE}
\int_{\phi(\gamma, t)} \boldsymbol{\sigma}^f(\boldsymbol{x}, t) \cdot \mathbf{n} \, \mathrm{d}\Gamma + \int_{\gamma} \boldsymbol{F}^s \, \mathrm{d}\Gamma = 0 \quad \forall \, \gamma \in \Gamma,
\end{equation}
where $\mathbf{n}$ is the outer normal to the interface.
To account for the evolving position of the fluid–structure interface, the motion of each spatial point within the fluid domain is explicitly controlled to maintain kinematic consistency at the discretized interface. A partitioned approach is used to couple the fluid and structural solvers. 
The effectiveness of the numerical model has been validated in previous studies that involved a range of fluid-structure interaction problems~\cite{jaiman2022computational, jaiman2016stable}. For further details on the numerical algorithm and implementation, interested readers may refer to~\cite{jaiman2022computational}.

\subsection{Discretization in Space and Time}
The governing equations~\ref{DNS}–\ref{INTERFACE} can be reformulated into the following abstract dynamical system:
\begin{equation}\label{DYNAMIC}
\frac{d\boldsymbol{q}}{dt} = \tilde{F}(\boldsymbol{q}),
\end{equation}
where \( \boldsymbol{q} \) represents the state of the coupled system, consisting of the fluid state, characterized by the velocity \( \boldsymbol{u}^f \) and pressure \( p \), and the solid state, which is implicitly captured through the mesh displacement. The function \( \tilde{F} \) denotes the governing dynamics of the coupled fluid–structure system. Upon temporal discretization with a fixed time step, Eq.~\ref{DYNAMIC} can be expressed using the forward Euler integration scheme:
\begin{equation}\label{eq:forwardEuler}
\boldsymbol{q}_{t_{n+1}} - \boldsymbol{q}_{t_n} = {F}(\boldsymbol{q}_{t_n}),
\end{equation}
where \( \boldsymbol{q}_{t_n} \) denotes the system state at the discrete time step \( t_n \), and \( F \) represents the full-order update function that governs the evolution of the system in the discretized time domain.
Equation~\ref{eq:forwardEuler} can then be spatially discretized on a computational mesh, resulting in:
\begin{equation}
	\label{eq:mesh}
	\boldsymbol{Q}_{t_{n+1}}-\boldsymbol{Q}_{t_n}=\boldsymbol{F}(\boldsymbol{Q}_{t_n}),
\end{equation}
where \( \boldsymbol{Q}_{t_n} \) denotes the system state variable matrix at time step \( t_n \), and \( \boldsymbol{F} \) represents the update function obtained after spatial and temporal discretizations. To model the dynamics of the discretized system, we adopt a hypergraph neural network framework that builds on the spatial and temporal discretization of the system states $\boldsymbol{Q}$. These states form the basis for constructing the input features of the model, as described in the following.

\subsection{Hypergraph Neural Network Framework}
Mathematically, a hypergraph \( \mathcal{H} = (\mathcal{V}, \mathcal{E}) \) consists of a vertex set \( \mathcal{V} \) and a hyperedge set \( \mathcal{E} \), where each hyperedge \( e \in \mathcal{E} \) is a subset of the vertex set and may include any number of nodes. In standard graphs, by comparison, each edge connects exactly two nodes. Graph-based models typically encode connectivity using an adjacency matrix \( \mathbf{A} \in \{0,1\}^{N \times N} \), where \( N = |\mathcal{V}| \) is the number of nodes and \( A_{ij} = 1 \) indicates a direct connection between nodes \( i \) and \( j \). Hypergraphs are more naturally represented using an incidence matrix \( \mathbf{H} \in \{0,1\}^{|\mathcal{V}| \times |\mathcal{E}|} \), where \( H(v, e) = 1 \) if the vertex \( v \) belongs to the hyperedge \( e \). This representation enables the modeling of higher-order interactions without decomposing them into pairwise relationships~\cite{feng2019hypergraphneuralnetworks}. Extensions of this formulation include directed hypergraphs, where hyperedges have designated source and target sets, weighted hypergraphs, which assign importance to nodes or hyperedges, and probabilistic hypergraphs, where incidence values lie in \([0, 1]\), reflecting uncertainty or partial association between nodes and hyperedges~\cite{dai2023hypergraph}.

\begin{figure*}
\centering
\begin{subfigure}{0.35\textwidth}\centering
    \includegraphics[width=.75\linewidth]{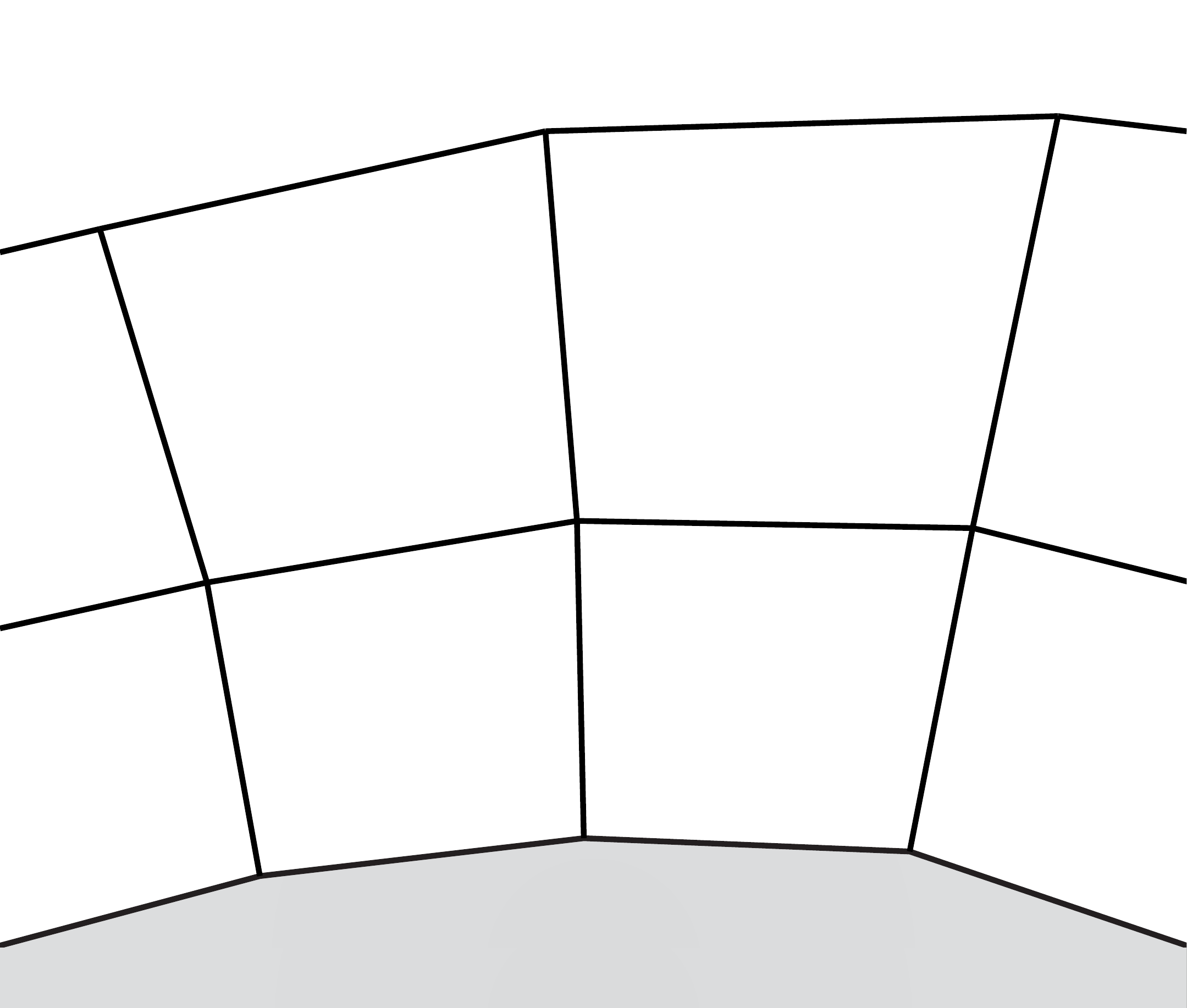}
    \caption{Computational grid}
\end{subfigure}
\begin{subfigure}{0.35\textwidth}\centering
    \includegraphics[width=.75\linewidth]{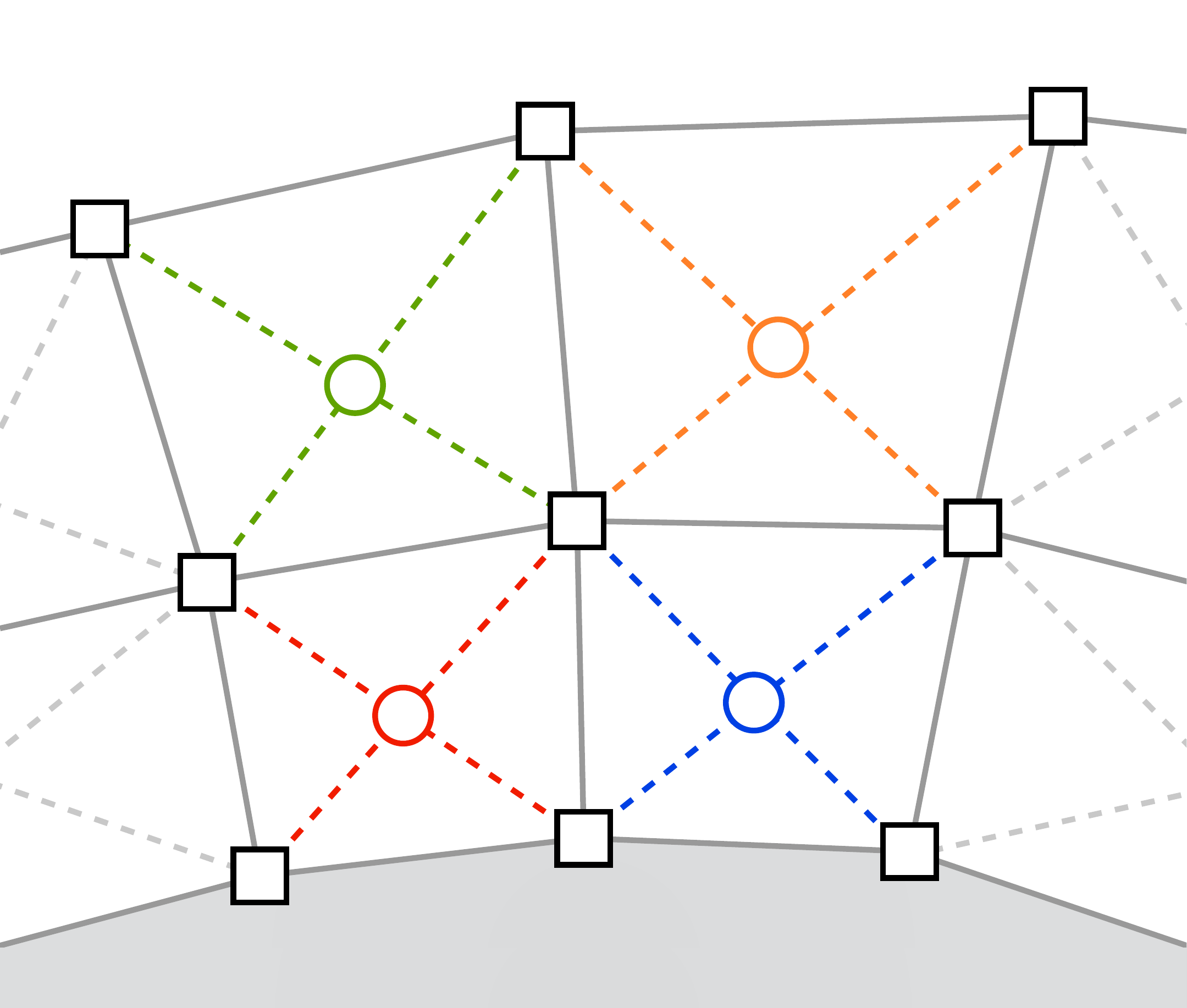}
    \caption{Node-element hypergraph}
\end{subfigure}
\caption{\label{cfd-gnn}
Transformation of a computational finite element mesh into a node-element hypergraph: (a) physical mesh representation, (b)  converted node-element hypergraph where each square ($\square$), circle ($\bigcirc$), and dashed line (- - -) represents a node, an element, and an element-node edge, respectively.}
\end{figure*}

\begin{figure*}
\centering\includegraphics[width=0.98\linewidth]{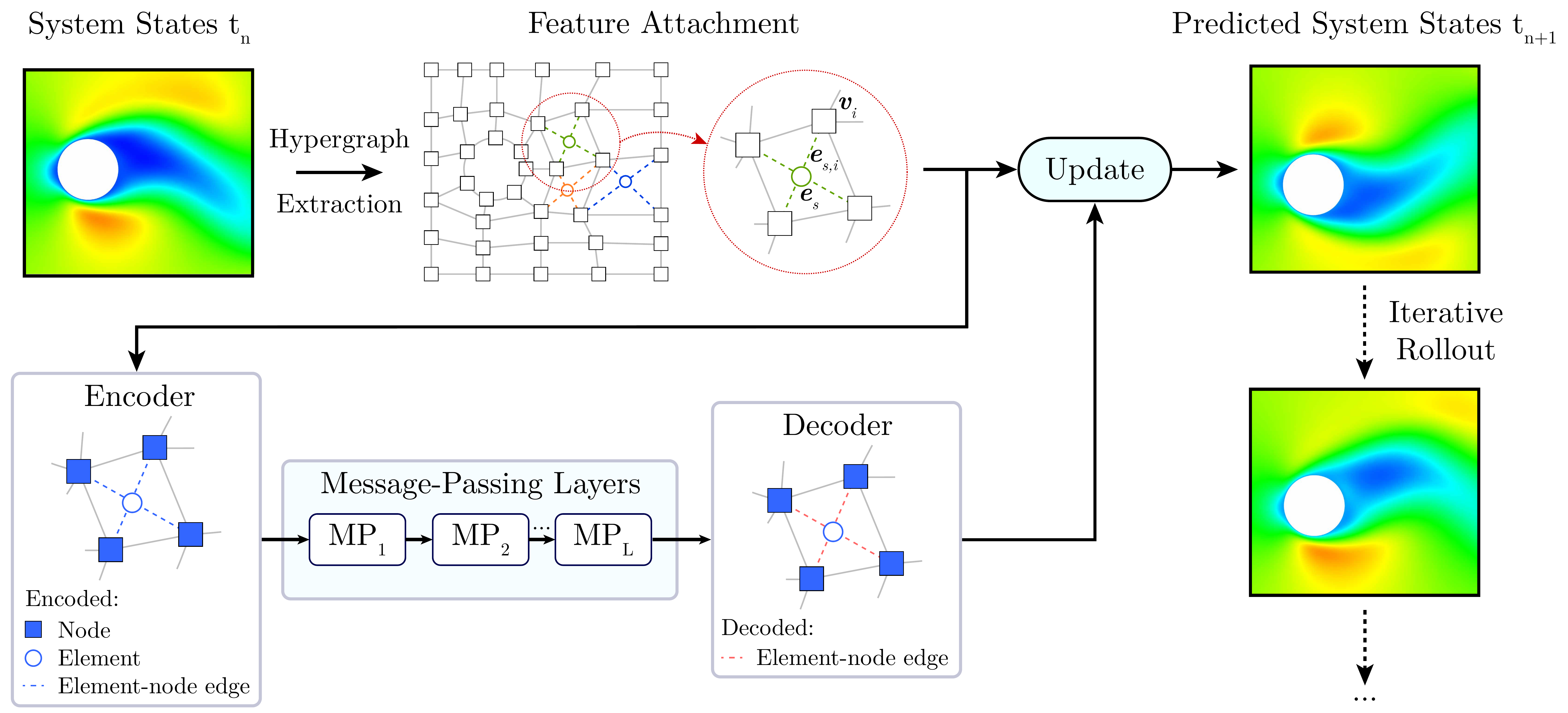}
\caption{\label{gnn_arch}Hypergraph neural network architecture for system states predictions during roll-out horizon.}
\end{figure*}

In the present study, we employ the recently developed finite element-inspired hypergraph neural network framework~\cite{gao2024finite, gao2024predicting}. This framework draws on the connectivity principles of finite element analysis by representing the computational mesh as a node-element hypergraph $\mathcal{G} = (\mathcal{V}, \mathcal{E}_s, \mathcal{E}_v)$, where $\mathcal{V}$ denotes the set of nodes, $\mathcal{E}_s$ the set of elements, and $\mathcal{E}_v$ the set of element–node edges, as illustrated in Fig.~\ref{cfd-gnn}. 
To construct the model inputs, a set of feature vectors is extracted from the system states $\boldsymbol{Q}$ and combined with geometric attributes of the computational mesh. These features are then embedded within the hypergraph structure. Specifically, we define the node feature vector $\boldsymbol{v}_i^{t_n}$ for each node $i$, the element feature vector $\boldsymbol{e}_s^{t_n}$ for each element $s$, and the element–node edge feature vector $\boldsymbol{e}_{s,i}^{t_n}$ for each edge connecting node $i$ to element $s$ at time step $t_n$, as shown in Fig.~\ref{gnn_arch}.
With these definitions, Eq.~\ref{eq:mesh} is reformulated as:
\begin{equation}
	\label{eq:graphstepfull}
	\boldsymbol{v}_i^{t_{n+1}}-\boldsymbol{v}_i^{t_{n}},\boldsymbol{e}_{s}^{t_{n+1}}-\boldsymbol{e}_{s}^{t_n},\boldsymbol{e}_{s,i}^{t_{n+1}}-\boldsymbol{e}_{s,i}^{t_n}=\boldsymbol{G}(\boldsymbol{v}_i^{t_{n}},\boldsymbol{e}_{s}^{t_n},\boldsymbol{e}_{s,i}^{t_n}),
\end{equation}
where $\boldsymbol{G}$ represents the hypergraph feature update function. A surrogate model based on a hypergraph neural network, denoted as \( \widehat{\boldsymbol{G}} \), is then constructed to approximate the feature update function \( \boldsymbol{G} \). 

A message-passing network is applied to the node-element hypergraph, emulating the computation of local stiffness matrices typically found in finite-element methods. The hypergraph neural network framework follows a three-step process: encoding, message-passing, and decoding, as seen in Fig.~\ref{gnn_arch}. The encoding step applied to the feature vectors is expressed as:
\begin{align}
    \boldsymbol{v}_i &\leftarrow g^v(\boldsymbol{v}_i), \quad 
    \boldsymbol{e}_{s} \leftarrow g^e(\boldsymbol{e}_{s}), \quad 
    \boldsymbol{e}_{s,i} \leftarrow g^{ev}(\boldsymbol{e}_{s,i}),
\end{align}
where the encoder functions \( g^v \), \( g^e \), and \( g^{ev} \) are implemented as multi-layer perceptrons (MLPs). The left arrow (\(\leftarrow\)) denotes an update operation, where the parameter on the left-hand side is assigned the computed value from the right-hand side. 

The feature vectors of encoded nodes, elements, and element-node edges are subsequently processed through a sequence of message-passing layers. Each layer comprises two stages: an element update stage and a node update stage. For an element \( s \) connected to four nodes \( i, j, k, \) and \( l \), the element update stage is formulated as:
\begin{equation}\label{eq:elem-upd}
    \boldsymbol{e}_{s} \leftarrow \text{AGG}_{r}^e \left( \phi^e (\boldsymbol{v}_r, \boldsymbol{e}_{s}, \boldsymbol{e}_{s,r}) \right), \quad
\end{equation}
where \(\text{AGG}\) represents an aggregation function, which is set to the mean function in the present work. The index \( r = i, j, k, l \) corresponds to the nodes connected to the element. The subsequent node update stage for each node \( i \) is given by:
\begin{equation}
    \boldsymbol{v}_i \leftarrow \text{AGG}_s^v \left( \phi^v (\boldsymbol{v}_i, \boldsymbol{e}^{'}_{s_i}, \boldsymbol{e}_{s_{i},i}) \right), \quad
\end{equation}
where, \( s_i \) represents any element that connects node \( i \) to other nodes, while \( \boldsymbol{e}^{'}_{s_i} \) denotes the element feature that has been updated in the previous element update stage (Eq.~\ref{eq:elem-upd}). The element and node update functions, \( \phi^e \) and \( \phi^v \), are not shared across different message-passing layers. At the end of the message-passing process, the outputs are decoded to generate the final predictions of the neural network. In this work, only the outputs associated with element-node edges are required, making an element-node decoder, \( h^{ev} \), sufficient for the task:
\begin{align}
    \boldsymbol{e}_{s,i} &\leftarrow h^{ev}(\boldsymbol{e}_{s,i}).
\end{align}
These outputs undergo additional post-processing to transform the results from element-node edges to node-level output vectors.

The encoding, element update, node update, and decoding functions are all implemented using multi-layer perceptrons. A multi-layer perceptron is a feedforward neural network that maps an input vector through multiple layers of linear transformations and nonlinear activation functions. This composition of transformations allows the network to extract complex hierarchical representations from the input data. For an input vector \( \boldsymbol{z} \), a multi-layer perceptron with \( L \) layers applies a sequence of functions, where each function corresponds to a layer transformation. The overall transformation can be expressed through function composition as:  
\begin{align}
\boldsymbol{h}(\boldsymbol{z}) = \boldsymbol{h}^{(L)} \circ \boldsymbol{h}^{(L-1)} \circ \dots \circ \boldsymbol{h}^{(1)} (\boldsymbol{z}),
\end{align}
where \( \circ \) denotes function composition, and each layer transformation \( \boldsymbol{h}^{(l)} \) is defined as:
\begin{align}
\boldsymbol{h}^{(l)} = \sigma^{(l)} \left( \boldsymbol{W}^{(l)} \boldsymbol{h}^{(l-1)} + \boldsymbol{b}^{(l)} \right), \quad l = 1, \dots, L-1.
\end{align}
In this formulation, \( \boldsymbol{h}^{(l)} \) represents the output of layer \( l \), while \( \boldsymbol{h}^{(0)} = \boldsymbol{z} \) is simply the input vector. The parameters \( \boldsymbol{W}^{(l)} \) and \( \boldsymbol{b}^{(l)}  \) define the trainable weight matrix and bias vector at each layer \( l \), respectively. The layer width is defined as the number of rows in the weight matrix. The function \( \sigma^{(l)}(\cdot) \) is a nonlinear activation function. In the final layer, the transformation takes the form:
\begin{align}
\boldsymbol{h}(\boldsymbol{z}) = \boldsymbol{W}^{(L)} \boldsymbol{h}^{(L-1)} + \boldsymbol{b}^{(L)},
\end{align}
where no activation function is applied. The composition of these transformations ensures that each intermediate representation is progressively refined, capturing both low-level and high-level features within the network. In the next subsection, we further expand on the framework architecture.

\subsection{Framework Details}
We examine the fluid-structure system on a bounded two-dimensional domain discretized using a body-fitting grid. At each vertex of the grid, the system state and corresponding vertex coordinates ($x$, $y$) are provided. Depending on the problem context, either the Reynolds number or the reduced velocity is assumed to be known. These values are then transformed into input features for the hypergraph neural network framework. The goal of this framework is to predict the flow variables $u_x$, $u_y$, and $p$ and the mesh displacements $\Delta x$ and $\Delta y$ at all vertices at each time step $t_{n+1}$ using the flow and mesh displacement information from previous time steps. To achieve this goal, we employ two distinct sub-networks: one dedicated to modeling the temporal evolution of the mesh state, and the other focused on predicting the flow state. Each sub-network is discussed separately in the following. While some details are similar to those in our earlier work \cite{gao2024predicting}, they are included here for completeness and ease of reference.

\subsubsection{Mesh State Prediction Sub-network}
For a two-dimensional grid, the mesh displacements \( \Delta\boldsymbol{x}_{t_n} \) and \( \Delta\boldsymbol{y}_{t_n} \) at each time step \( t_n \) are combined into a single complex-valued vector given by:
\begin{equation}
  \Delta\boldsymbol{{{X}}}_{t_n} \;=\; \Delta \boldsymbol{x}_{t_n} \;+\; \mathrm{i}\, \Delta \boldsymbol{y}_{t_n},
\end{equation}
where $\mathrm{i}$ denotes the unit imaginary number. Vertex displacements are computed relative to their initial positions, ensuring that the mesh state predictions remain invariant under domain translations. The resulting complex-valued vectors from the training set are then assembled into a matrix $\Delta \boldsymbol{X} \in \mathbb{C}^{P \times N}$, where $P$ is the number of grid vertices and $N$ is the total number of time snapshots. Then a singular value decomposition is applied to $\Delta \boldsymbol{X}$, which yields:
\begin{equation}
  \Delta \boldsymbol{X} \;=\; \boldsymbol{U}\,\boldsymbol{\Sigma}\,\boldsymbol{V}^H.
\end{equation}
Here, the superscript \(H\) indicates the conjugate transpose operation. The complex-valued proper orthogonal decomposition is used to enforce rotation equivariance in network predictions (see \cite{gao2024predicting} for details). The singular values located on the diagonal of \(\boldsymbol{\Sigma}\) are arranged in descending order, and the leading vectors (columns) of the matrix \(\boldsymbol{U}\) are termed modes. Assuming the dataset has a low-rank structure, a reduced-order model can be developed by projecting \(\Delta \boldsymbol{X}\) onto the first \(m\) dominant modes. Mathematically, this projection is expressed as:
\begin{align}
  \boldsymbol{C}^{(1:m)} 
  \;=\; \bigl(\boldsymbol{U}^{(1:m)}\bigr)^H \,{\Delta \boldsymbol{X}}, \\
    \Delta \widehat{\boldsymbol{X}} 
  \;=\; \boldsymbol{U}^{(1:m)} \,\boldsymbol{C}^{(1:m)},
\end{align}
where \(\boldsymbol{U}^{(1:m)}\) represents the matrix \(\boldsymbol{U}\) truncated to its first \(m\) columns, thereby preserving only the leading \(m\) modes. The matrix \(\boldsymbol{C}^{(1:m)}\) contains the projection coefficients corresponding to each retained mode, and \(\Delta \widehat{\boldsymbol{X}}\) denotes the reduced-order approximation of \(\Delta \boldsymbol{X}\) based on these \(m\) dominant modes. Consequently, the dynamics of the original system is effectively described by the evolution of these \(m\) mode coefficients over time.
Based on the above formulation, the approximate mesh displacement vector \(\Delta \widehat{\boldsymbol{X}}_{t_n}\) at each time step \(t_n\) is computed as:
\begin{align}\label{eq:POD-1}
    \Delta \widehat{\boldsymbol{X}}_{t_n}
  \;=\; \boldsymbol{U}^{(1:m)} \,\boldsymbol{C}_{t_n}^{(1:m)},
\end{align}
where the coefficients $\boldsymbol{C}_{t_n}^{(1:m)}$ at each time step $t_n$ are given by:
\begin{equation}\label{eq:POD-2}
  \boldsymbol{C}_{t_n}^{(1:m)} \;=\; \boldsymbol{\rho}_{t_n}^{(1:m)}\, \ \mathrm{exp} \ \bigr( \mathrm{i} \,\boldsymbol{\eta}_{t_n}^{(1:m)}\bigl).
\end{equation}
We utilize multi-layer perceptrons to predict the coefficients \(\boldsymbol{C}_{t_{n+1}}^{(1:m)}\) at time step \(t_{n+1}\). For this prediction, the coefficients from the preceding \(h\) time steps, specifically \(t_n, t_{n-1}, \dots, t_{n-h+1}\), are assumed to be known. Using these known coefficients, the increments of \(\boldsymbol{\rho}^{(1:m)}\) and \(\boldsymbol{\eta}^{(1:m)}\) are first calculated between consecutive time steps within this history window. For example, the increments at time step \(t_{n}\) are calculated as:
\begin{align}
    \delta \boldsymbol{\rho}_{t_n}^{(1:m)} = \boldsymbol{\rho}_{t_n}^{(1:m)} - \boldsymbol{\rho}_{t_{n-1}}^{(1:m)}, \\
\delta \boldsymbol{\eta}_{t_n}^{(1:m)} = \boldsymbol{\eta}_{t_n}^{(1:m)} - \boldsymbol{\eta}_{t_{n-1}}^{(1:m)}.
\end{align}
The resulting increment sequences \(\delta \boldsymbol{\rho}_{t_n}^{(1:m)}, \delta \boldsymbol{\rho}_{t_{n-1}}^{(1:m)}, \dots, \delta \boldsymbol{\rho}_{t_{n-h+1}}^{(1:m)}\), along with the increment sequences \(\delta \boldsymbol{\eta}_{t_n}^{(1:m)}, \delta \boldsymbol{\eta}_{t_{n-1}}^{(1:m)}, \dots, \delta \boldsymbol{\eta}_{t_{n-h+1}}^{(1:m)}\), are then provided as inputs to the multi-layer perceptrons to predict the increments \(\delta \tilde{\boldsymbol{\rho}}_{t_{n+1}}^{(1:m)}\) and \(\delta \tilde{\boldsymbol{\eta}}_{t_{n+1}}^{(1:m)}\) at time step $t_{n+1}$. Finally, a forward Euler scheme is employed to update the coefficients as:
\begin{align}
    \tilde{\boldsymbol{\rho}}_{t_{n+1}}^{(1:m)} = \tilde{\boldsymbol{\rho}}_{t_n}^{(1:m)} + \delta \tilde{\boldsymbol{\rho}}_{t_{n+1}}^{(1:m)}, \\
\tilde{\boldsymbol{\eta}}_{t_{n+1}}^{(1:m)} = \tilde{\boldsymbol{\eta}}_{t_n}^{(1:m)} + \delta \tilde{\boldsymbol{\eta}}_{t_{n+1}}^{(1:m)}.
\end{align}
Based on the formulations provided in Eqs.~\ref{eq:POD-1} and \ref{eq:POD-2}, the predicted mesh displacement \(\Delta \tilde{\boldsymbol{X}}_{t_{n+1}}\) at the next time step \(t_{n+1}\) is computed as follows:
\begin{align}
    \Delta \Tilde{\boldsymbol{X}}_{t_{n+1}}
  \;=\; \boldsymbol{U}^{(1:m)} \,\Tilde{\boldsymbol{{\rho}}}_{t_{n+1}}^{(1:m)}\, \ \mathrm{exp} \ \bigr( \mathrm{i} \,\Tilde{\boldsymbol{{\eta}}}_{t_{n+1}}^{(1:m)}\bigl),
\end{align}
where the real and imaginary parts correspond to the $x$ and $y$ components of the mesh displacement, respectively. 
The adopted autoregressive approach offers a conceptually and computationally straightforward implementation, as it enables reuse of a single-step prediction architecture without requiring complex sequence encoding or decoding mechanisms. In addition, it aligns well with real-time prediction and control applications, since it does not rely on access to the full temporal context and supports incremental deployment of predictions.
The complex-valued POD enables rotation-equivariant representations of mesh displacement, particularly critical for capturing motion trajectories in flow-induced vibration. The phase and amplitude separation allows the model to naturally encode the rotational symmetry and cyclic behavior of structural oscillations, which would otherwise require a larger latent space in a real-valued setting.

\subsubsection{Flow State Prediction Sub-network}
The sub-network for the temporal propagation of the flow state employs a forward Euler scheme that iteratively predicts feature vector increments between successive time steps. Initially, the finite element computational grid is converted into a node-element hypergraph. This hypergraph undergoes encoding, message-passing, and decoding stages, as shown in Fig.~\ref{gnn_arch}, wherein multi-layer perceptrons function as encoders, decoders, and update mechanisms within the sub-network. 
Subsequently, the predicted features are post-processed to derive the physical flow states at each node. The boundary condition vector $\boldsymbol{\kappa}_i$ is assigned to each node feature \( \boldsymbol{v}_i \), i.e., $\boldsymbol{v}_i = \boldsymbol{\kappa}_i$. The area of each element $A_s$ is taken as the element feature. Instead of directly using the area, we utilize its natural logarithm to account for variations spanning several orders of magnitude between large and small elements. After further feature augmentation, the resulting element feature vector is defined as:
\begin{align}
\boldsymbol{e}_{s} = 
\begin{Bmatrix}
\ln A_{s} &
-\ln A_{s}
\end{Bmatrix}^T.
\end{align}

To define the element-node edge feature \( \boldsymbol{e}_{s,i} \), we first perform initial transformations to extract geometric, mesh, and flow features. For an element $s$ connecting nodes $i, j, k,$ and $l$, the element center coordinates $(x_{s}, y_{s})$ are computed by averaging the coordinates $x$ and $y$ coordinates over nodes $r = i, j, k, l$. The local coordinates of each node relative to the element center can then be expressed as:
\begin{align}
    \boldsymbol{{x}}_{s,r} = (x_{s,r}, y_{s,r}) = (x_r -{x_s},\, y_r - {y_s}).
\end{align}

The lengths of these local coordinate vectors, denoted by $L_{s,r}$, along with the angles at the corners of the element $\theta_{s,r}$ and the element area $A_{s}$, constitute the geometric features. Together, they fully define the shape and size of the element.
For the mesh feature, the mesh displacement $(\Delta x_r, \Delta y_r)$ is projected onto the directions defined by the unit local coordinate vector $\boldsymbol{\hat{x}}_{s,r}$ associated with each node $r$ to obtain the following:
\begin{align}
d_{s,r} 
= 
\frac{1}{L_{s,r}}
\begin{Bmatrix}
\Delta x_r & \Delta y_r
\end{Bmatrix}
\begin{Bmatrix}
x_{s,r} \\
y_{s,r}
\end{Bmatrix},
\end{align}
where $d_{s,r}$ represents the weights of the mesh displacement projections. For the flow feature, the velocity vector $(u_{x,r},u_{y,r})$ is projected onto the direction of the unit local coordinate vector $\boldsymbol{\hat{x}}_{s,r}$, following the same approach used for the mesh displacement:
\begin{align}
f_{s,r} 
= 
\frac{1}{L_{s,r}}
\begin{Bmatrix}
u_{x,r} & u_{y,r}
\end{Bmatrix}
\begin{Bmatrix}
x_{s,r} \\
y_{s,r}
\end{Bmatrix},
\end{align}
where $f_{s,r}$ represents the weights of the flow velocity projections. The scalar pressure value \( p_r \) at each node \( r \) is directly assigned to the corresponding element-node edge:
\begin{align}
    p_{s,r} = p_r.
\end{align}
The signed square root of \( p_{s,r} \) is then computed and concatenated with the available features to construct the element-node edge feature vector \( \boldsymbol{e}_{s,r} \), defined as:
\begin{align}
\boldsymbol{e}_{s,r} = 
\begin{Bmatrix}
f_{s,r} & 
\mathrm{sgn}(p_{s,r})\sqrt{|p_{s,r}|} &
d_{s,r} &
K^* &
\ln L_{s,r} &
\cos \theta_{s,r}
\end{Bmatrix}^T.
\end{align}
Here, \( \mathrm{sgn}(\cdot) \) represents the sign function and \( K^* \) is a dimensionless parameter, either the Reynolds number or the reduced velocity, chosen based on the specific problem.

Using the forward Euler scheme, the network iteratively predicts the increments of the feature vectors between successive time steps. The hypergraph neural network then updates the projected velocity components and the associated pressure values encoded in the edge feature vectors of the element node at each step. These updated features are then post-processed to recover the physical flow variables, namely the flow velocity \((u_x, u_y)\) and pressure \(p\), at each node. The scalar pressure at a given node is obtained by averaging the predicted values across all element-node edges connected to that node. The original values are reconstructed for vector quantities by reversing the projection using the Moore–Penrose pseudo-inverse~\cite{10.1063/5.0097679}. In the following, we present the results obtained using the hypergraph neural network framework to predict flow-induced vibrations in isolated and tandem cylinders.

\section{Results and Discussion}
\label{sec:results}
We begin by describing the procedure used to generate ground truth data, followed by the network architecture and training process. The predictive performance of the hypergraph neural network framework is then evaluated by comparing its roll-out predictions with ground-truth data for the isolated and tandem configurations.

\subsection{Ground Truth Data Generation}
\begin{figure*}
\centering
\begin{subfigure}{0.75\textwidth}\centering
    \includegraphics[width=1\linewidth]{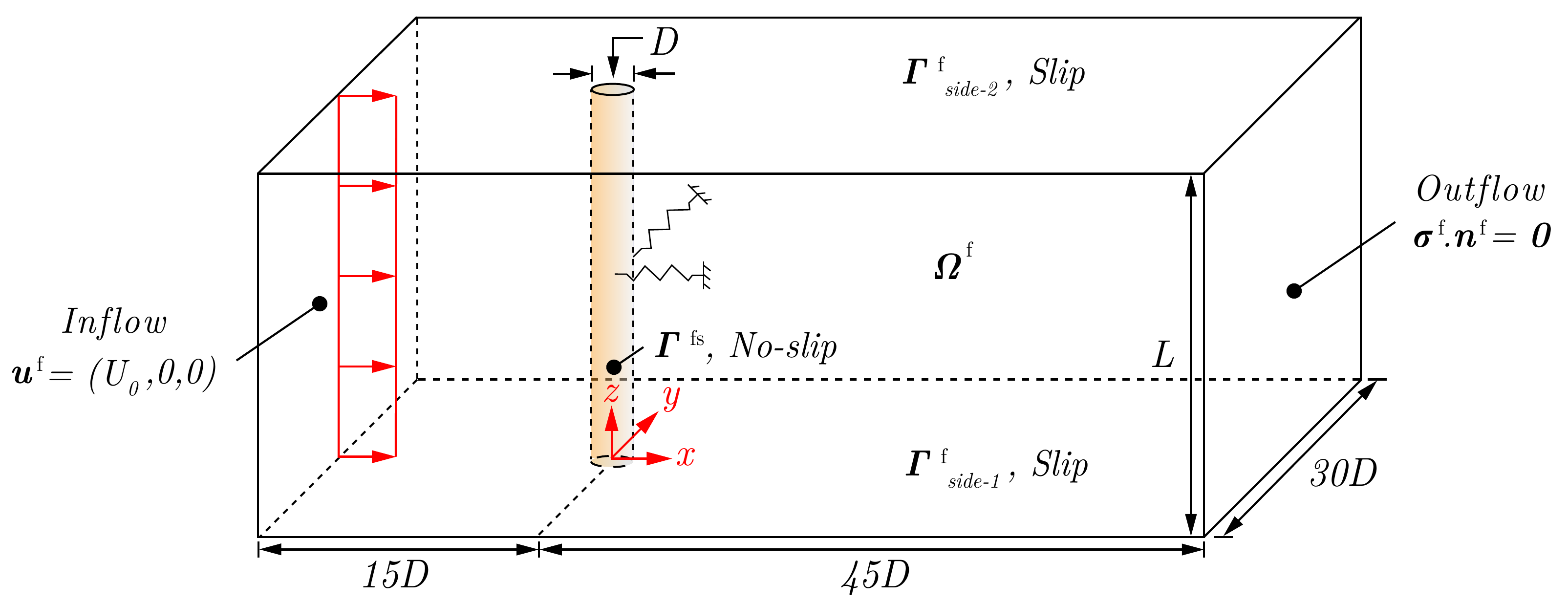}
    \caption{}
\end{subfigure}\\
\begin{subfigure}{0.8\textwidth}\centering
    \includegraphics[width=1\linewidth]{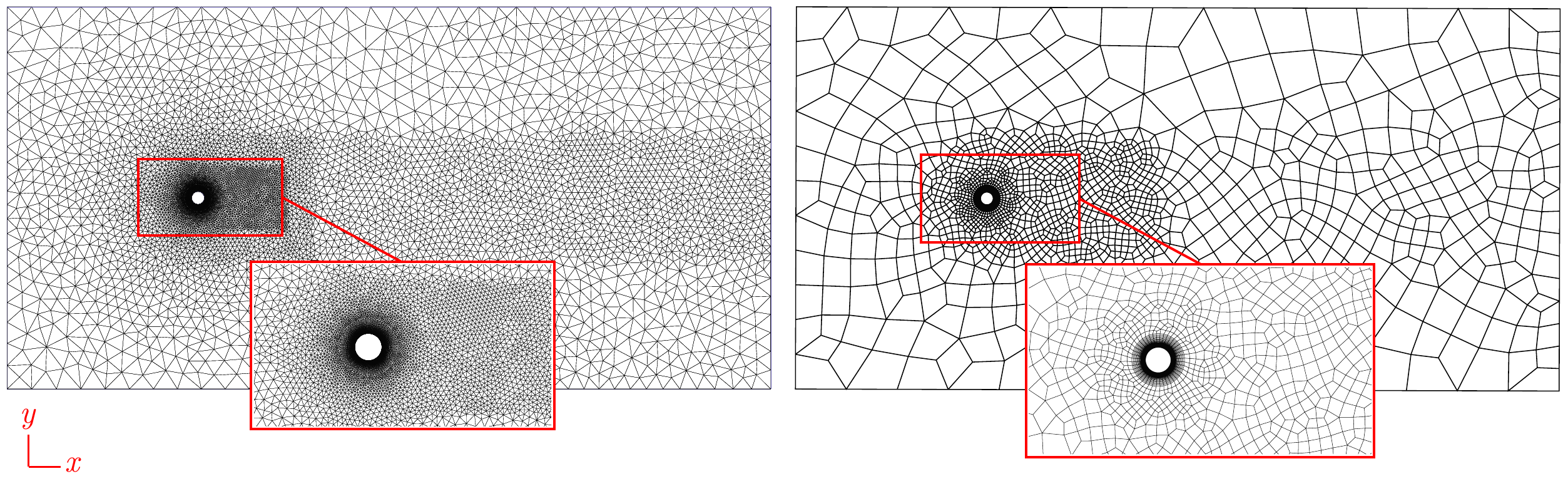}\\
    \caption{}
\end{subfigure}
\caption{Computational domain and mesh setup for isolated configuration. (a) Schematic of the computational domain used for full-order simulations, along with the imposed boundary conditions. (b) The computational meshes used for full-order and hypergraph neural network modeling; (left) the mesh for the ground truth full-order simulation and (right) the mesh for the hypergraph neural network framework. For the ground truth simulation, only a single layer of the mesh is used in the \(z\)-direction, so the \(x\)-\(y\) plane configuration is presented. The neural network mesh is two-dimensional.}
\label{fig:domain_iso}
\end{figure*}

\begin{figure*}
\centering
\begin{subfigure}{0.75\textwidth}\centering
    \includegraphics[width=1\linewidth]{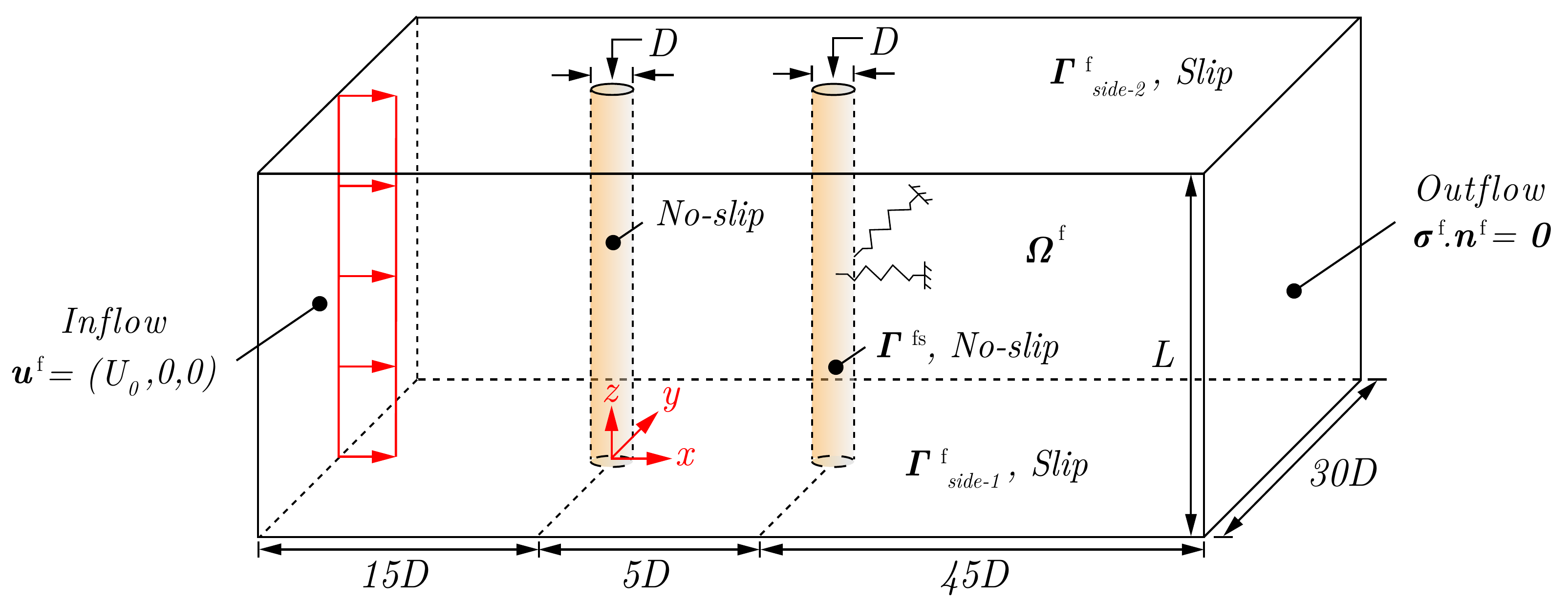}
    \caption{}
\end{subfigure}\\
\begin{subfigure}{0.8\textwidth}\centering
    \includegraphics[width=1\linewidth]{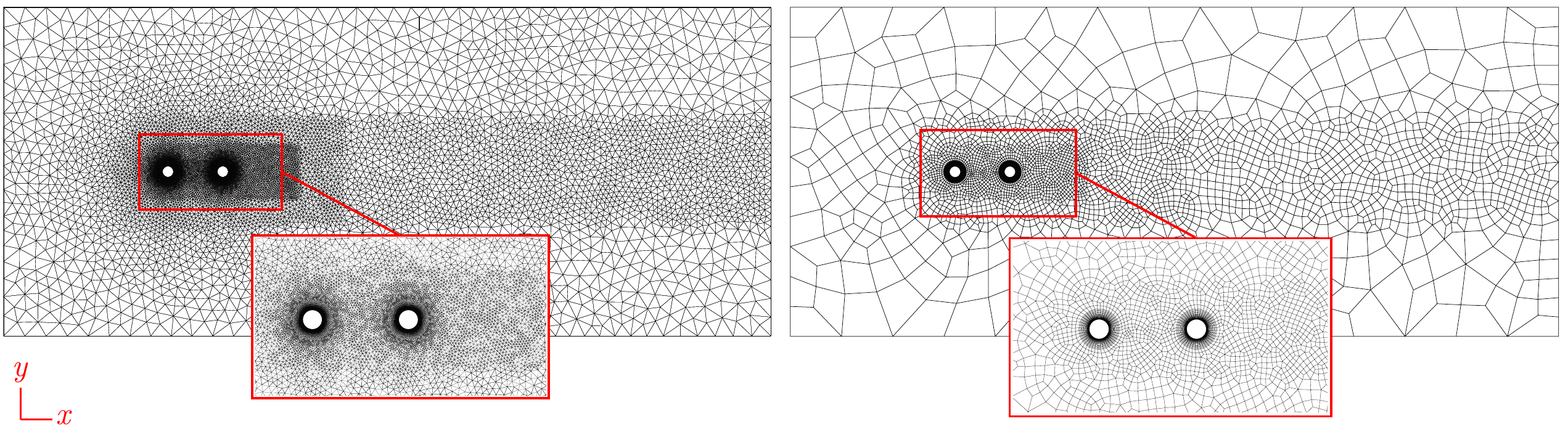}\\
    \caption{}
\end{subfigure}
\caption{Computational domain and mesh setup for tandem configuration. (a) Schematic of the computational domain used for full-order simulations, along with the imposed boundary conditions. (b) The computational meshes used for full-order and hypergraph neural network modeling; (left) the mesh for the ground truth full-order simulation and (right) the mesh for the hypergraph neural network framework.}
\label{fig:domain-tandem}
\end{figure*}

Building upon the formulations presented in the previous section, the ground-truth data for training and evaluating the framework are obtained from high-fidelity full-order simulations. The simulation domain, along with the boundary conditions, used to generate the ground truth data is shown in Fig. \ref{fig:domain_iso}a for the isolated configuration. 
A no-slip boundary condition is enforced on the surface of the oscillating cylinder, denoted by \(\Gamma^{fs}\). At the inflow boundary, a uniform flow of velocity \(\boldsymbol{u}^f = (U_0, 0, 0)\) is prescribed, where $U_0$ is the magnitude of the free-stream flow velocity. Slip boundary conditions are applied on the side surfaces of the simulation domain. At the outflow boundary, a traction-free condition is imposed, expressed as \( \boldsymbol{\sigma}^f \cdot \boldsymbol{n}^f = \boldsymbol{0} \), where \( \boldsymbol{n}^f \) denotes the unit normal vector to the outflow surface. In addition, a periodic boundary condition is applied in the spanwise \(z\) direction.  
For the tandem configuration, the computational domain and boundary conditions are depicted in Fig.~\ref{fig:domain-tandem}a. The boundary conditions are consistent with those of the isolated case, with the addition of a no-slip condition applied to the surface of the stationary upstream cylinder.

For each configuration, the ground truth data are recorded once the system reaches a statistically stationary state. For the isolated and tandem configurations, the training dataset includes 2176 and 3072 consecutive time steps, respectively, beginning from time step $h + 1$, where $h = 7$ denotes the number of historical steps used by the mesh prediction sub-network. To evaluate the performance of the hypergraph neural network, test data comprising 1001 consecutive time steps are collected, also starting from the time step $h + 1$.
To reduce computational overhead, the data are interpolated from the full-order simulation mesh onto a coarser mesh, shown in Figs.~\ref{fig:domain_iso}b and ~\ref{fig:domain-tandem}b for the isolated and tandem configurations, respectively, using the Clough-Tocher 2D interpolation scheme from the SciPy package \cite{virtanen2020scipy}. The interpolated data are then split into training and testing sets for use in the hypergraph neural network framework.

The non-dimensional parameters considered in this work include the mass ratio ($m^\mathrm{*}$), Reynolds number $Re$, and reduced velocity $U^*$ given by:
\begin{align}
	m^\mathrm{*} = \frac{4 m^s}{{\pi D^{2}\rho^{f}}}, \qquad
	Re = \frac{\rho^{f}U_{0}D}{\mu^{f}}, \qquad
	U^\mathrm{*} &= \frac{U_0}{f_{s}D},
\end{align}
where $m^s$ is the cylinder's mass per unit length, $D$ is the cylinder diameter, $\rho^{f}$ is the fluid density, $\mu^{f}$ is the dynamic viscosity of the fluid, and $f_{s}$ is the cylinder's natural frequency. All simulations are conducted with a mass ratio of $m^* = 1$ and zero structural damping. For the isolated configuration, the Reynolds number is fixed at $Re = 60$, and the reduced velocity $U^*$ varies from 3.5 to 7.0 in increments of 0.25, resulting in 15 cases used for training. To evaluate the framework, test data are generated at reduced velocities $U^* = 4.125$, $5.125$, and $6.125$. A constant time step size of $\Delta t = 0.001$ is used for all simulations in this configuration. For the tandem arrangement, the reduced velocity is fixed at $U^* = 9$. The Reynolds number ranges from $45$ to $100$ in increments of 2.5, yielding 23 cases. Among these, cases at $Re = 50, 60, 70, 80$, and $90$ are reserved for testing, while the remaining are used for training. The time step size is set to $\Delta t = 0.003$ for simulations in the tandem configuration. The number of retained modes is fixed at $m = 2$ across all cases in isolated and tandem arrangements.

To verify that the interpolated mesh accurately captures the system dynamics, we compare the lift and drag coefficients computed on the interpolated mesh with those obtained from the reference simulation mesh. The lift and drag coefficients are defined as follows:
\begin{align}
	C_{l} = \frac{1}{\frac{1}{2}\rho^{f}U_\mathrm{0}^\mathrm{2}D}\int_{\Gamma^{fs}} (\boldsymbol{\sigma}^{f}\cdot \boldsymbol{n})\cdot \boldsymbol{n}_{y} \mathrm{d\Gamma}, \quad \\
C_{d} = \frac{1}{\frac{1}{2}\rho^{f}U_\mathrm{0}^\mathrm{2}D}\int_{\Gamma^{fs}} (\boldsymbol{\sigma}^{f}\cdot \boldsymbol{n})\cdot \boldsymbol{n}_{x} \mathrm{d\Gamma}, \quad
\end{align}
where $C_{l}$ and $C_{d}$ denote the lift and drag coefficients, respectively, and $\boldsymbol{n}_{x}$ and $\boldsymbol{n}_{y}$ are the Cartesian components of the unit outward normal vector $\boldsymbol{n}$.
Table~\ref{tab:meshInfo} summarizes this comparison for the isolated configuration. As shown in Table~\ref{tab:meshInfo}, the relative errors in the lift and drag coefficients remain below 1\%, confirming the accuracy of the interpolated mesh. 
\begin{table*}
\caption{Comparison of mesh properties and hydrodynamic coefficients for the interpolated and reference meshes at $Re = 60$, $m^* = 1$, and $U^* = 5$ in isolated cylinder configuration. The values in parentheses indicate relative errors compared to the reference mesh results.}
\centering
\begin{tabular}{lcc}
\hline
 & \textrm{Interpolated mesh} & \textrm{Reference mesh} \\
\hline
Element type & Quadrilateral & Wedge \\
Number of nodes & 3184 & 18256 \\
Number of elements & 3133 & 18058 \\
Mean drag coefficient $\overline{C_{d}}$ & 2.1158 (0.53\%) & 2.1270 \\
root mean square lift coefficient $C_{l}^{rms}$ & 0.4167 (0.81\%) & 0.4201  \\
\hline
\end{tabular}
\label{tab:meshInfo}
\end{table*}

\subsection{Network and Training Details}
The framework uses the PyTorch library~\cite{NEURIPS2019_9015}. The mesh and flow prediction sub-networks are trained independently, using a single NVIDIA A100 GPU for the isolated configuration and an NVIDIA V100 GPU for the tandem configuration. Min-max normalization with a range of \([-1, 1]\) is applied to each entry of input and output features in the training datasets, excluding \(\cos \theta_{s,i}\) in the element-node edge features, as its values are inherently bounded within \((-1, 1)\). The test datasets are normalized using the statistics computed from the training datasets. The mesh prediction sub-network, trained with a batch size of 256, incorporates a Lookahead optimization scheme~\cite{zhang2019lookahead} wrapped around a base AdamW optimizer~\cite{loshchilov2019}. The AdamW optimizer is configured with a learning rate of \(10^{-2}\) and a weight decay of \(10^{-3}\). The Lookahead optimization refines the base optimizer by updating the weights every \(k = 5\) steps using a slow-moving average with an interpolation factor \(\alpha = 0.5\). This combination allows the model to benefit from the fast adaptation of AdamW while leveraging the stabilizing effects of Lookahead’s long-term weight updates. 

Multi-layer perceptrons, consisting of 4 hidden layers, each with a width of 512 neurons, are used for the mesh displacement prediction. The sinusoidal activation function is applied across the sub-network layers, along with the corresponding initialization scheme~\cite{sitzmann2020implicit}. The loss function for the mesh prediction sub-network is defined as a weighted sum of the adaptive smooth $L_1$ loss and the log-cosh loss. The hybrid loss is formulated as:
\begin{align}
\mathcal{L}_i &= \lambda_1 \cdot \mathcal{L}_i^{{L_1}} + \lambda_2 \cdot \log\left( \cosh(\hat{\psi}_i - \psi_i) + \varepsilon \right),
\end{align}
where the adaptive smooth $L_1$ loss is given by:
\begin{align}
\mathcal{L}_i^{{L_1}} =
\begin{cases}
\dfrac{1}{2\beta} (\hat{\psi}_i - \psi_i)^2, & \text{if } |\hat{\psi}_i - \psi_i| < \beta, \\
|\hat{\psi}_i - \psi_i| - \dfrac{\beta}{2}, & \text{otherwise}.
\end{cases}
\end{align}
Here, $\hat{\psi}_i$ and $\psi_i$ denote the predicted and ground truth values, respectively. The subscript $i$ denotes entry-by-entry computation over all prediction–target pairs. The parameter $\beta$ is adapted online during training based on the exponential moving average of the residual variance. A constant $\varepsilon = 10^{-6}$ is added to the logarithm for numerical stability. The weighting coefficients are set to $\lambda_1 = 0.8$ and $\lambda_2 = 0.2$. 

A hybrid cosine-to-exponential scheduler is utilized for the mesh prediction sub-network~\cite{gao2024finite}, starting the learning rate from \(10^{-4}\) and gradually decreasing it to \(10^{-6}\) over 350 epochs. To prevent exploding gradients, gradient clipping~\cite{pascanu2013difficulty} is applied to the parameters of the mesh prediction sub-network. A clipping threshold of 0.1 is specified such that if the total gradient norm exceeds the threshold, the gradients are proportionally scaled to maintain stability. During training of the mesh prediction sub-network, zero-mean Gaussian noise with a standard deviation of $\sigma = 6 \times 10^{-3}$ is added to the input and output values of the ground truth data to promote robustness in predictions across simulation cases. To monitor performance, the mesh prediction sub-network is trained using a validation subset composed of the final 15\% of temporally ordered samples from each training case. Early stopping is applied with a patience of 80 epochs, stopping training if the validation loss does not improve by at least $10^{-6}$ during this period. The model corresponding to the lowest validation loss is retained for evaluation. 
%
%

The flow prediction sub-network, trained with a batch size of 4, uses the Adam optimizer~\cite{kingma2017}, with $\beta_1 = 0.9$ and $\beta_2 =  0.999$. Following the approach in \cite{gao2024finite}, a hybrid cosine-to-exponential scheduler is employed for this sub-network. The learning rate schedule spans 50 training epochs, beginning with a 5-epoch warm-up period during which the learning rate increases per iteration from $10^{-6}$ to $10^{-4}$. Following warm-up, the learning rate decreases from $10^{-4}$ to $10^{-6}$ over the remaining 45 epochs, with updates applied per epoch according to a cosine-to-exponential schedule.

We use multi-layer perceptrons for the encoders, the decoder, and the node and element update functions within each network layer. Each multi-layer perceptron contains two hidden layers and uses a sinusoidal activation function. All multi-layer perceptrons in the hypergraph neural network have a fixed layer width of 128 neurons. The architecture includes 15 message-passing layers based on an internal ablation study. Increasing depth from 8 to 15 layers improved accuracy in pressure field prediction and wake recovery, while further increases yielded diminishing returns and risked over-smoothing. This depth ensures sufficient receptive field coverage for multi-element interactions in wake-structured flows.
As with the mesh prediction sub-network, Gaussian noise is injected during the training of the flow prediction sub-network, using \(\sigma = 10^{-1}\) and an over-correction factor of 1.2. An adaptive smooth \(L_1\) function is employed as the loss function for the flow prediction sub-network. 

A summary of the sub-network architectures and training configurations is provided in Table~\ref{tab:network_summary}. The mesh prediction sub-network requires approximately 1 second per training epoch, while the flow prediction sub-network requires approximately 750 seconds per epoch for the isolated configuration and 2770 seconds per epoch for the tandem configuration. After the training is complete, the roll-out predictions over 1000 time steps, starting from time step \(h+1\), are generated and then compared against the ground truth values. Inference was conducted on CPUs paired with the respective GPU configurations available during training. For the isolated configuration, an AMD EPYC 7413 (Zen 3) operating at 2.65 GHz was employed, yielding an average inference time of approximately 10 ms per time step. In the tandem configuration, an Intel Xeon Silver 4216 (Cascade Lake) running at 2.1 GHz resulted in an inference time of approximately 30 ms per time step. In the following, we present the prediction results.
\begin{table}
\centering
\caption{Summary of architecture and training settings for the mesh and flow state prediction sub-networks.}
\label{tab:network_summary}
\begin{tabular}{ll}
\hline
\textbf{Component} & \textbf{Details} \\
\hline
\textbf{Mesh Prediction Sub-network} & \\
\quad Optimizer & Lookahead (update every 5 steps, $\alpha=0.5$) + AdamW \\
\quad Learning rate schedule & Cosine-to-exponential decay ($10^{-4} \rightarrow 10^{-6}$, 350 epochs) \\
\quad Gradient clipping & Global norm threshold of 0.1 \\
\quad Noise injection & Gaussian noise ($\sigma = 6 \times 10^{-3}$) \\
\quad Network architecture & 4-layer MLP (512 neurons per layer) \\
\quad Batch size & 256 \\
\hline
\textbf{Flow Prediction Sub-network} & \\
\quad Optimizer & Adam \\
\quad Learning rate schedule & Cosine-to-exponential decay ($10^{-4} \rightarrow 10^{-6}$, 50 epochs) \\
\quad Noise injection & Gaussian noise ($\sigma = 10^{-1}$) \\
\quad Network architecture & 15-layer message-passing hypergraph network; \\
& 2-layer MLP (128 neurons per layer) \\
\quad Batch size & 4 \\
\hline
\end{tabular}
\end{table}

\subsection{System State Predictions in Isolated Configuration}
\begin{figure*}
\centering
\begin{subfigure}{0.8\textwidth}\centering
    \includegraphics[width=1\linewidth]{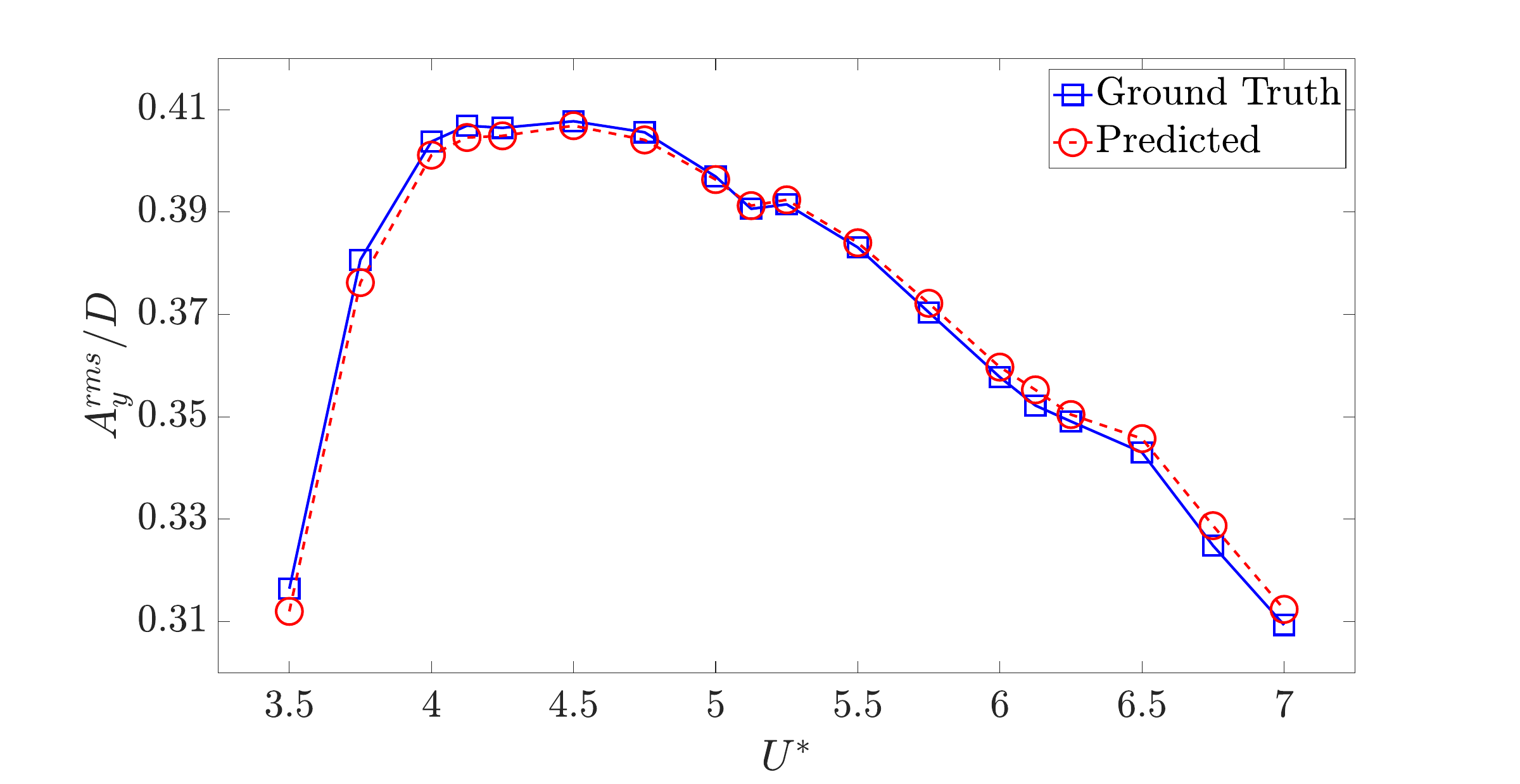}
    \caption{}
\end{subfigure}
\begin{subfigure}{0.49\textwidth}\centering
    \includegraphics[width=1\linewidth]{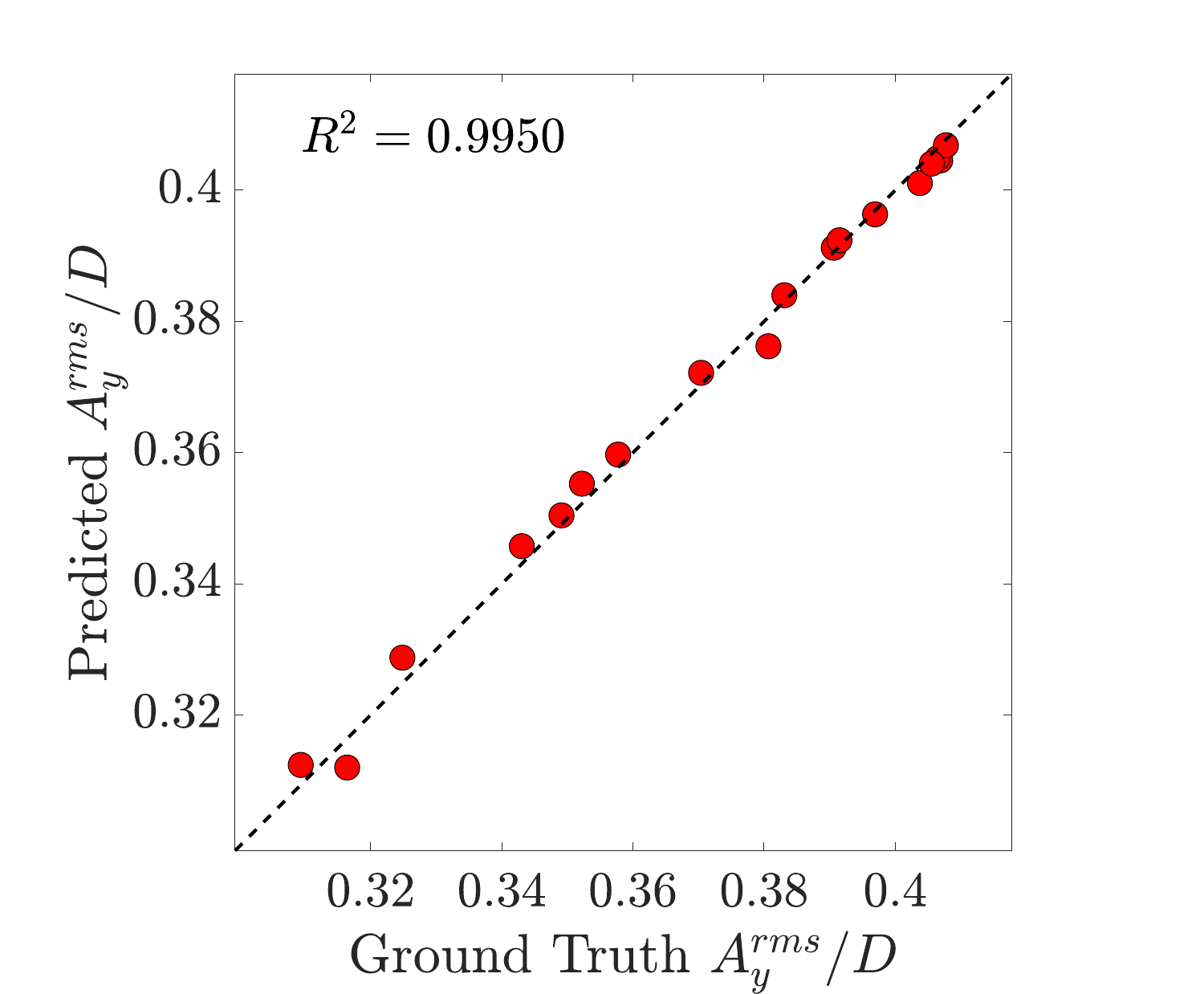}
    \caption{}
\end{subfigure}
\begin{subfigure}{0.49\textwidth}\centering
    \includegraphics[width=1\linewidth]{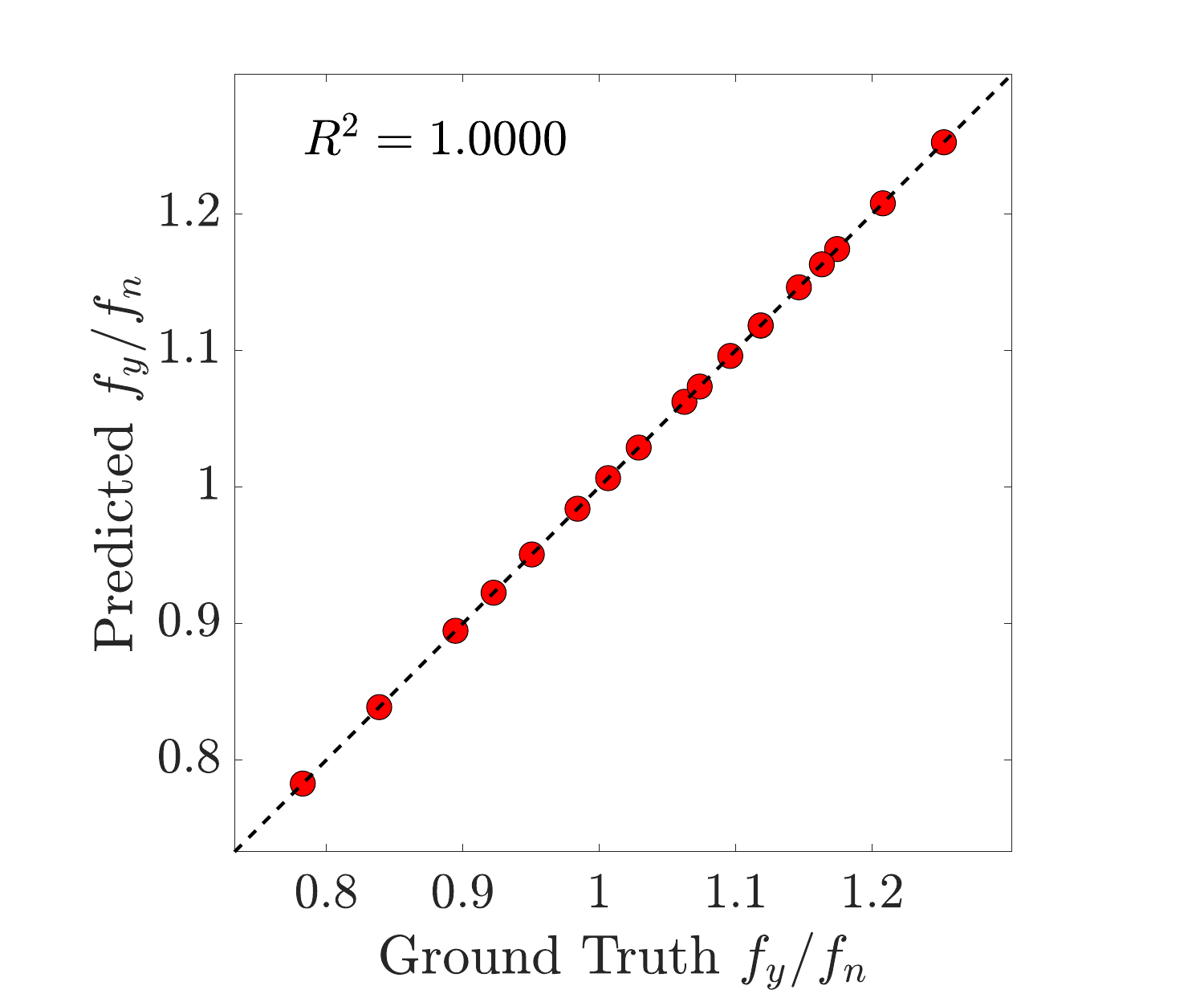}
    \caption{}
\end{subfigure}
\caption{\label{Ay-fy-Re60-VIV}Model performance in predicting transverse oscillation amplitude and frequency in isolated configuration. (a) Comparison of the transverse oscillation amplitude as a function of reduced velocity $U^*$ between the ground truth results and data obtained using the trained model. (b) Comparison between predicted and ground truth values for the normalized root mean square fluctuating transverse vibration amplitude \(A_y^{{rms}}/D\) and (c) the dominant transverse oscillation frequency \(f_y/f_n\) across training and test datasets.}
\end{figure*}

A primary objective in the prediction of FIVs is to accurately capture the variation of the oscillation amplitude with respect to system parameters. To this end, we compare the root mean square value of the normalized fluctuating transverse vibration amplitude (\(A_y^{rms}/D\)) obtained using the trained mesh prediction sub-network with ground truth data obtained from high-fidelity simulations. Figure~\ref{Ay-fy-Re60-VIV}a presents this comparison across training and test datasets. The trained model demonstrates good agreement with the ground truth, successfully capturing the onset of large-amplitude oscillations around \(U^* \approx 3.5\), the peak amplitude near \(U^* \approx 4\), and the gradual decline in amplitude at higher values of \(U^*\). Thus, capturing the characteristic bell-shaped amplitude trend with respect to \( U^* \). The coefficient of determination (\(R^2\)), computed as the square of the Pearson correlation coefficient between the predicted and ground truth values, is used to quantitatively assess the model’s predictive ability. 
Figures~\ref{Ay-fy-Re60-VIV}b and \ref{Ay-fy-Re60-VIV}c present a comparison between the predicted and ground truth values of the normalized oscillation amplitude $A_y^{{rms}}/D$ and the dominant transverse oscillation frequency ($f_y/f_n$), respectively, for the isolated configuration at $Re = 60$. Here, $f_n$ denotes the cylinder’s natural frequency, accounting for the added mass effect due to the surrounding fluid. The model achieves an excellent match for oscillation amplitudes, with a coefficient of determination of $R^2 = 0.9950$. For the dominant frequency prediction, the model yields a perfect agreement with the ground truth, achieving $R^2 = 1.0000$.

\begin{figure*}
\centering
\begin{subfigure}{0.49\textwidth}\centering
    \includegraphics[width=1\linewidth]{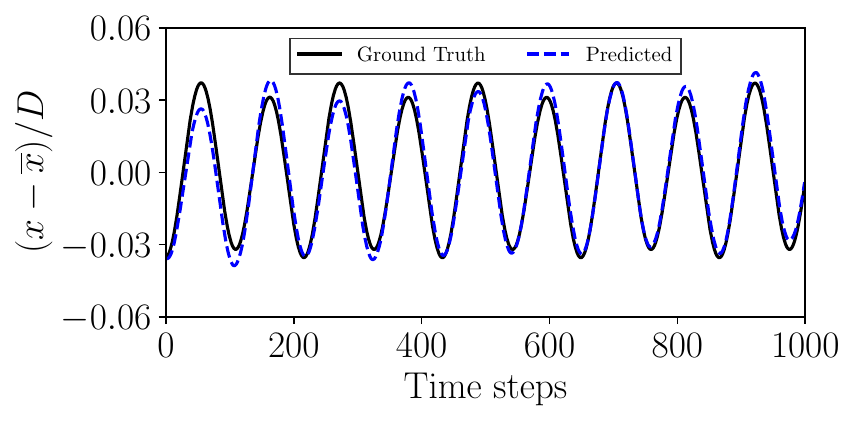}
    \caption{}
\end{subfigure}
\begin{subfigure}{0.27\textwidth}\centering
    \includegraphics[width=1\linewidth]{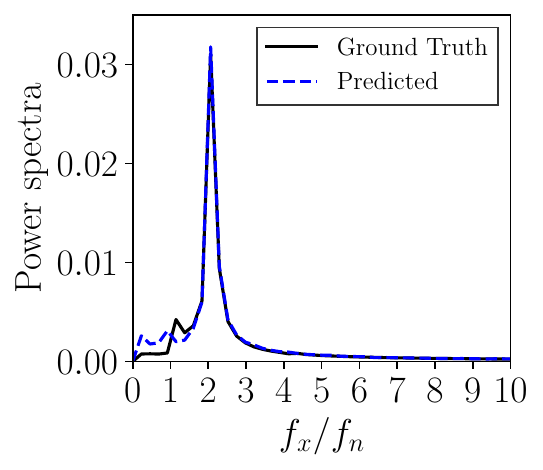}\\
    \caption{}
\end{subfigure}
\begin{subfigure}{0.49\textwidth}\centering
    \includegraphics[width=1\linewidth]{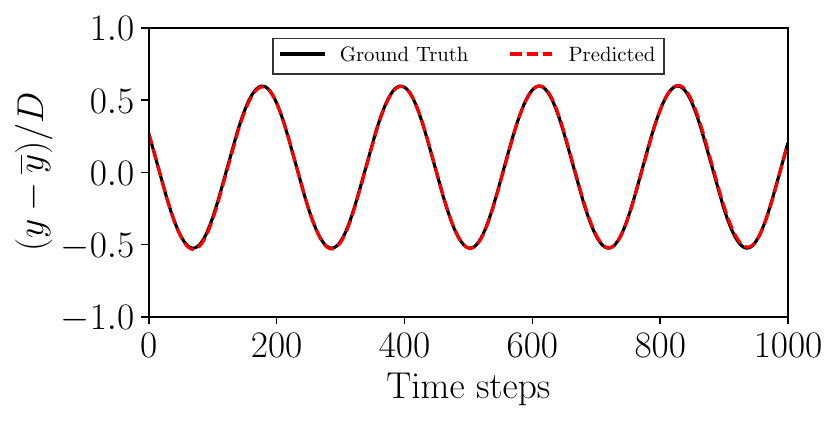}
    \caption{}
\end{subfigure}
\begin{subfigure}{0.27\textwidth}\centering
    \includegraphics[width=1\linewidth]{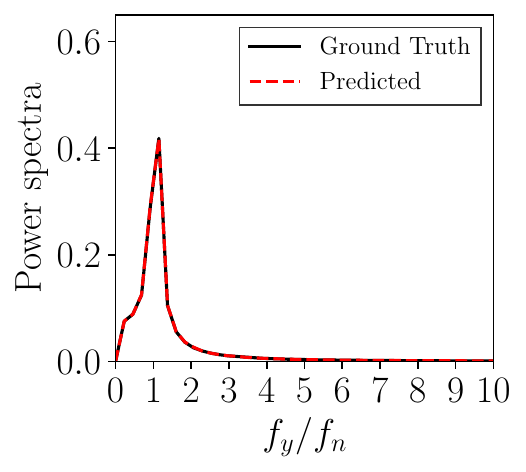}\\
    \caption{}
\end{subfigure}
\caption{\label{time_histories}Comparison of the cylinder displacement predictions with ground truth data in isolated configuration at a test \(U^* = 5.125\). Subplots (a) and (c) show the time histories of the normalized streamwise and transverse displacements, respectively. Subplots (b) and (d) present the corresponding power spectra.}
\end{figure*}

\begin{figure*}
\centering
\begin{subfigure}{0.32\textwidth}\centering
    \includegraphics[width=0.98\linewidth]{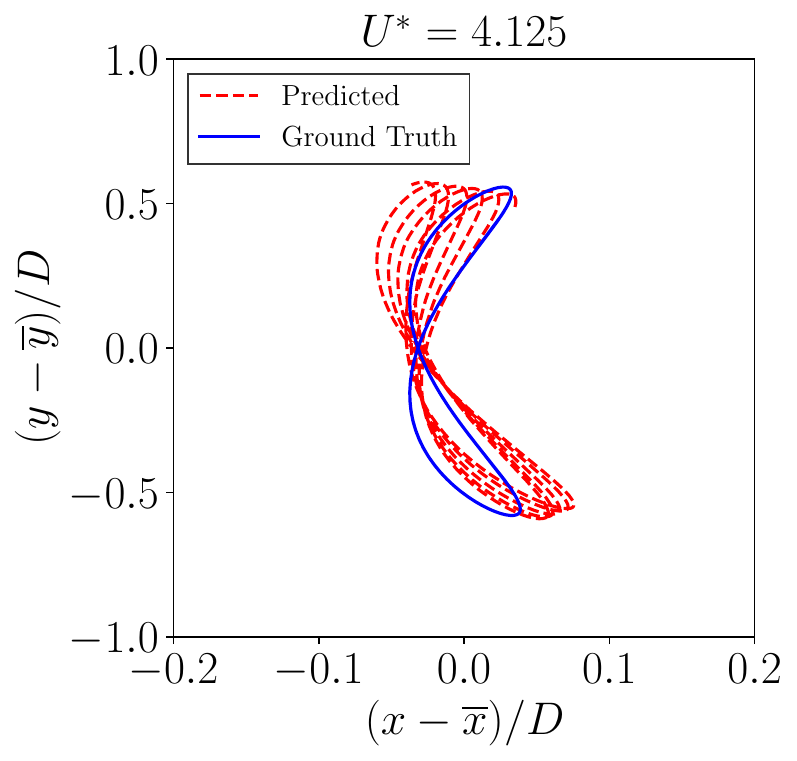}
    \caption{}
\end{subfigure}
\begin{subfigure}{0.32\textwidth}\centering
    \includegraphics[width=0.98\linewidth]{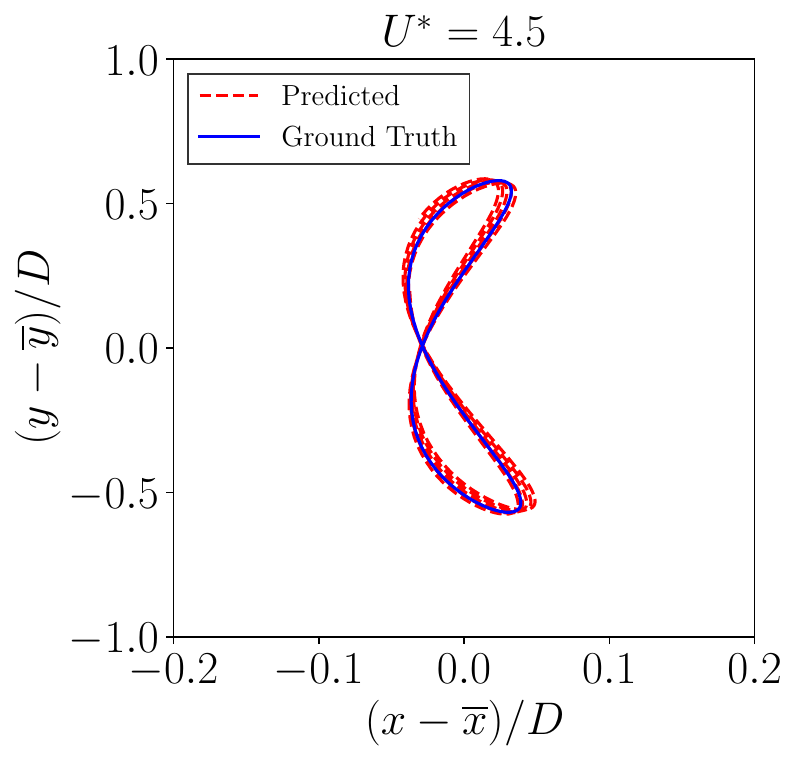}
    \caption{}
\end{subfigure}
\begin{subfigure}{0.32\textwidth}\centering
    \includegraphics[width=0.98\linewidth]{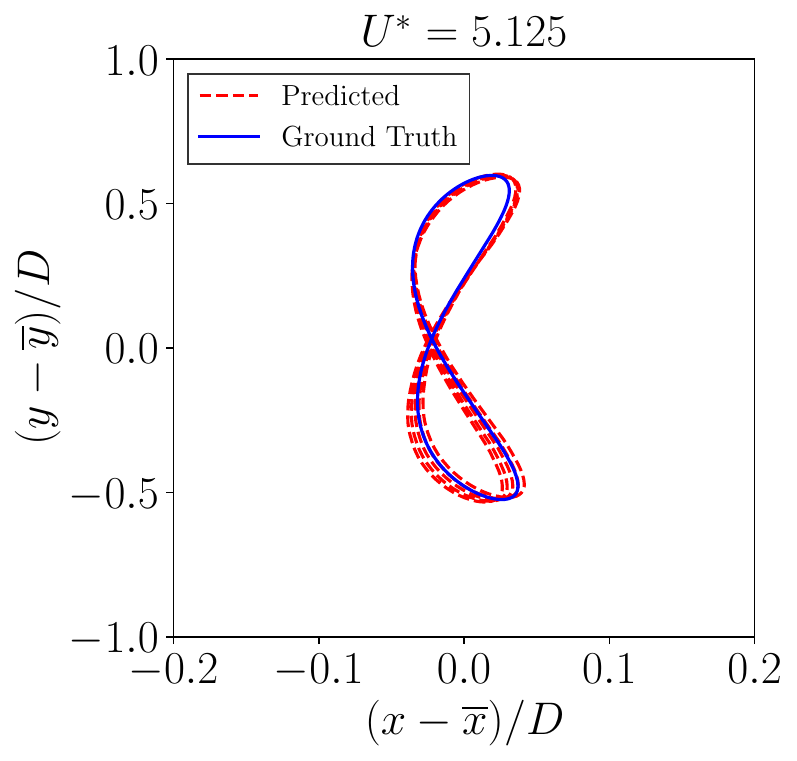}
    \caption{}
\end{subfigure}\\
\begin{subfigure}{0.32\textwidth}\centering
    \includegraphics[width=0.98\linewidth]{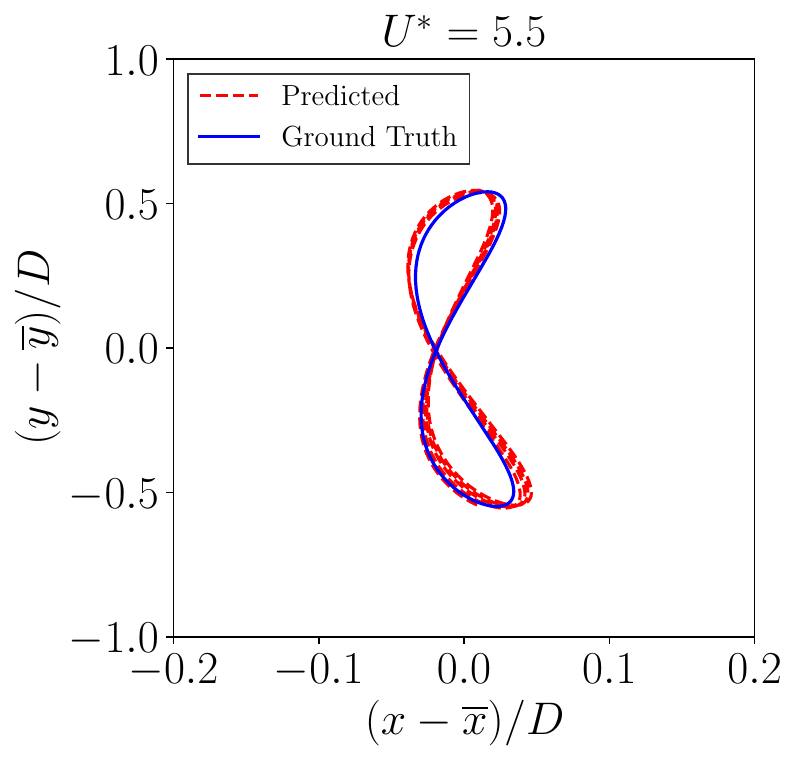}
    \caption{}
\end{subfigure}
\begin{subfigure}{0.32\textwidth}\centering
    \includegraphics[width=0.98\linewidth]{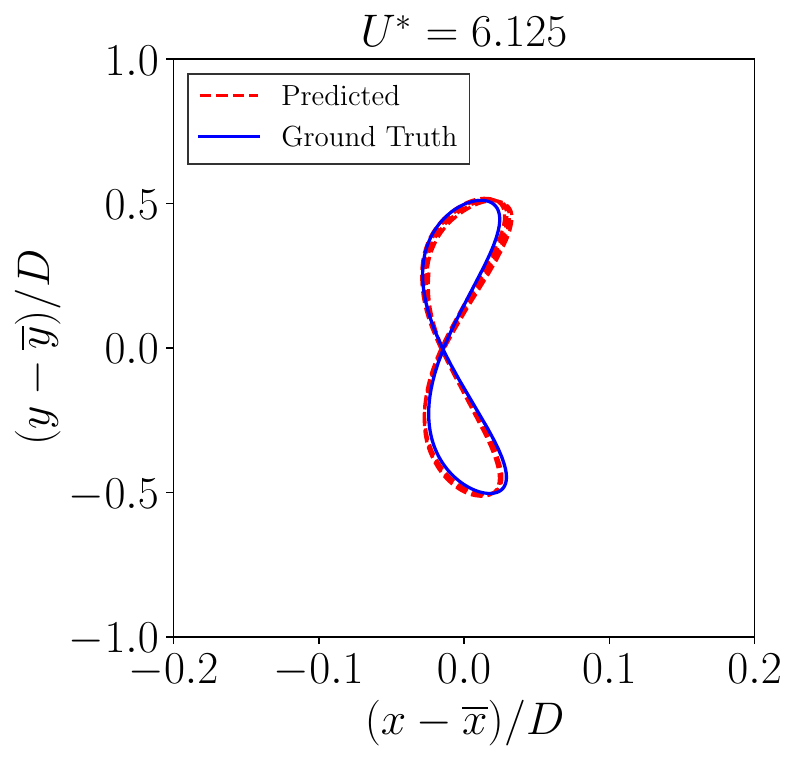}
    \caption{}
\end{subfigure}
\begin{subfigure}{0.32\textwidth}\centering
    \includegraphics[width=0.98\linewidth]{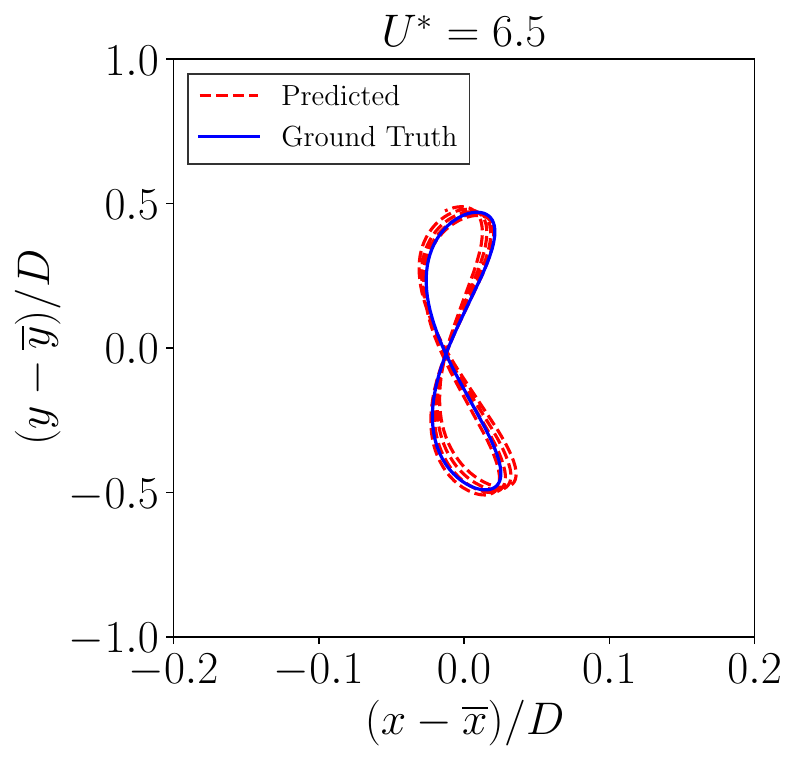}
    \caption{}
\end{subfigure}
\caption{\label{trajectories}Comparison of cylinder's motion trajectories between predictions and ground truth results for selected cases in isolated configuration.}
\end{figure*}

Figure~\ref{time_histories} presents the time histories of the cylinder's normalized streamwise and transverse displacements, denoted by \((x - \bar{x})/D\) and \((y - \bar{y})/D\), respectively, along with the corresponding power spectra of the oscillatory response, at a test reduced velocity of \(U^* = 5.125\). The model captures the oscillatory dynamics with high accuracy, as demonstrated by the close agreement between the predicted and ground truth values shown in Fig.~\ref{time_histories}. In the streamwise direction, the predicted values closely match the ground truth, with minor deviations observed near the peak values.
The motion trajectories in the \(x\)-\(y\) plane are shown in Fig.~\ref{trajectories} at selected cases. The model is found to accurately reconstruct the characteristic figure-eight patterns observed in the ground truth trajectories. Deviations in the streamwise displacement are observed, for instance at $U^* = 4.125$, where the predicted motion slightly diverges from the ground truth. These deviations are likely due to the relatively smaller amplitude of the streamwise motion, making it more sensitive to modeling errors. However, the errors have limited influence on the overall trajectory pattern, as the global motion structure is largely preserved.

\begin{figure*}
\centering
\begin{subfigure}{0.328\textwidth}\centering
    \includegraphics[width=1\linewidth]{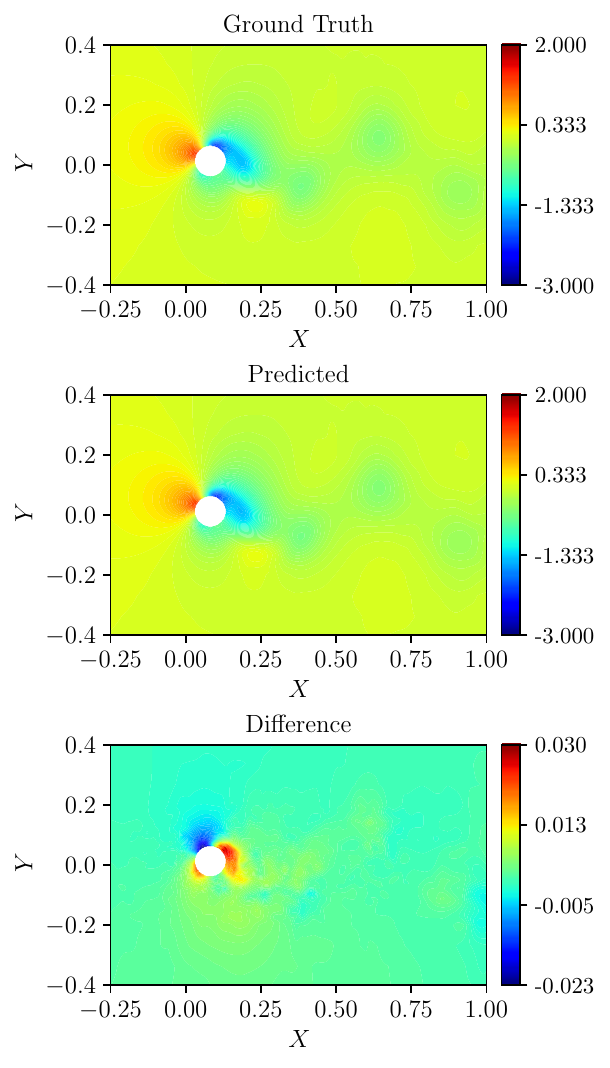}
    \caption{}
\end{subfigure}
\begin{subfigure}{0.328\textwidth}\centering
    \includegraphics[width=1\linewidth]{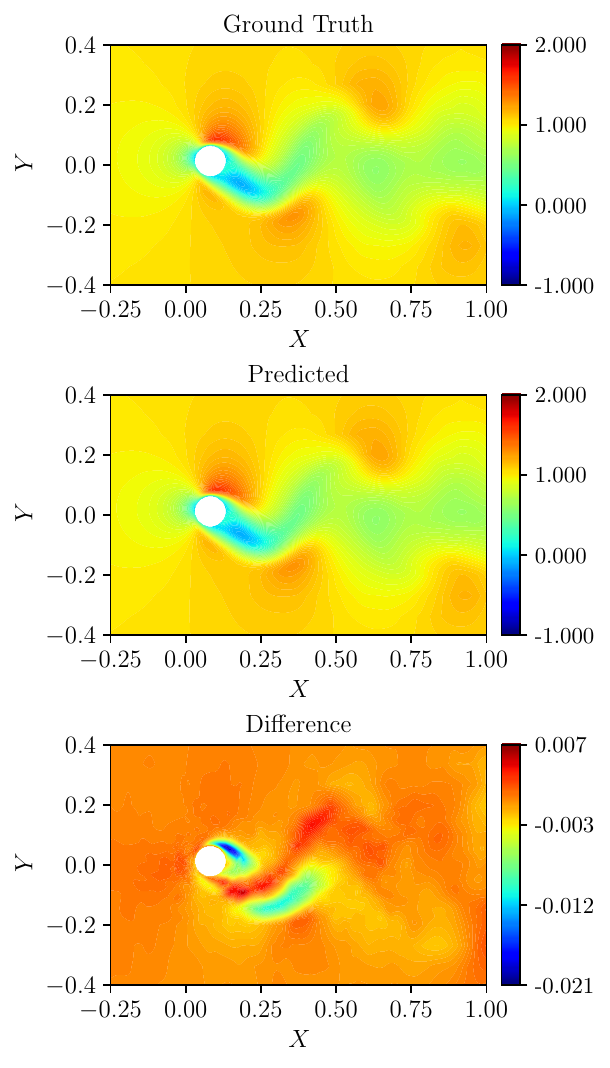}
    \caption{}
\end{subfigure}
\begin{subfigure}{0.328\textwidth}\centering
    \includegraphics[width=1\linewidth]{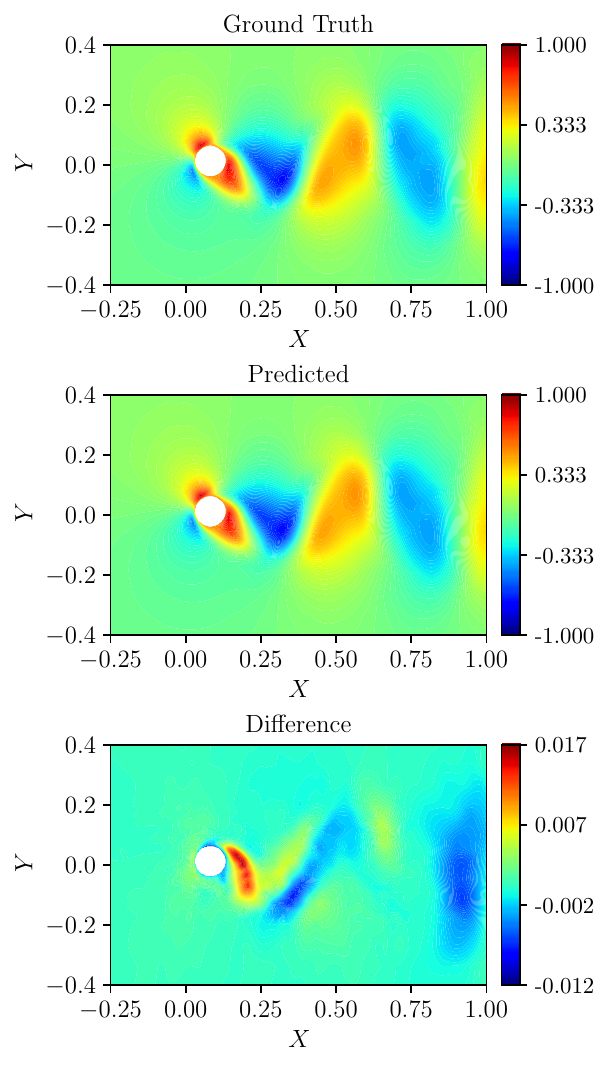}\\
    \caption{}
\end{subfigure}
\caption{\label{contours-VIV}Comparison between predicted and ground truth contours of normalized (a) pressure ($2p/\rho^{f}U_{0}^2$), (b) streamwise velocity ($u_{x}/U_{0}$), and (c) crossflow velocity ($u_{y}/U_{0}$) for the isolated cylinder at a test \(U^{*}=5.125\); fields are shown at roll‑out step 1000.}
\end{figure*}

\begin{figure*}
\centering
\begin{subfigure}{0.495\textwidth}\centering
    \includegraphics[width=1\linewidth]{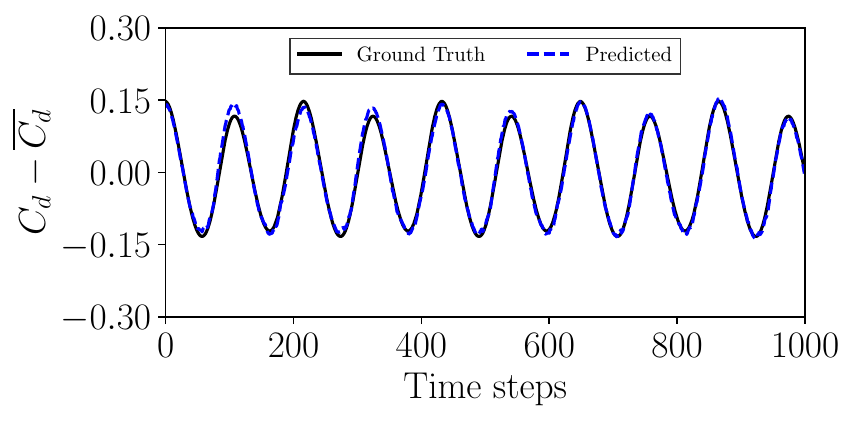}
    \caption{}
\end{subfigure}
\begin{subfigure}{0.275\textwidth}\centering
    \includegraphics[width=1\linewidth]{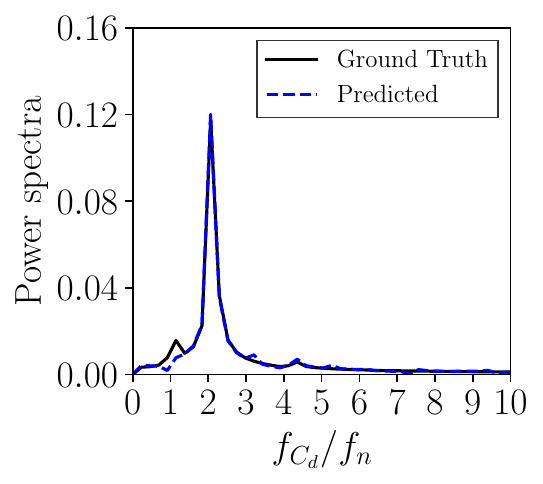}\\
    \caption{}
\end{subfigure}
\begin{subfigure}{0.49\textwidth}\centering
    \includegraphics[width=1\linewidth]{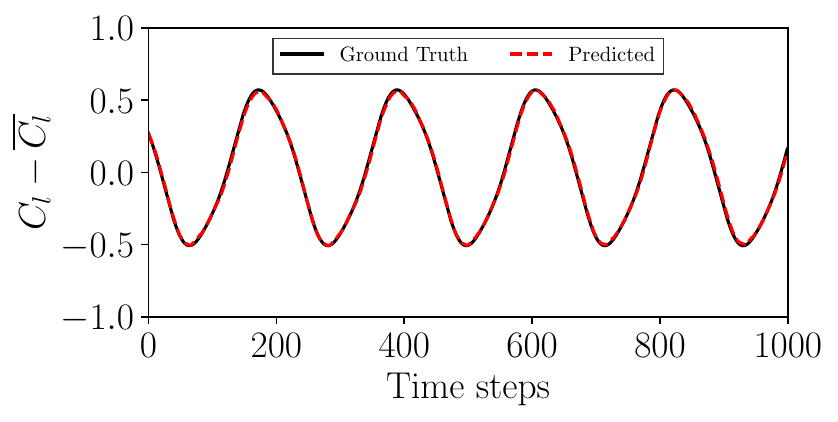}
    \caption{}
\end{subfigure}
\begin{subfigure}{0.275\textwidth}\centering
    \includegraphics[width=1\linewidth]{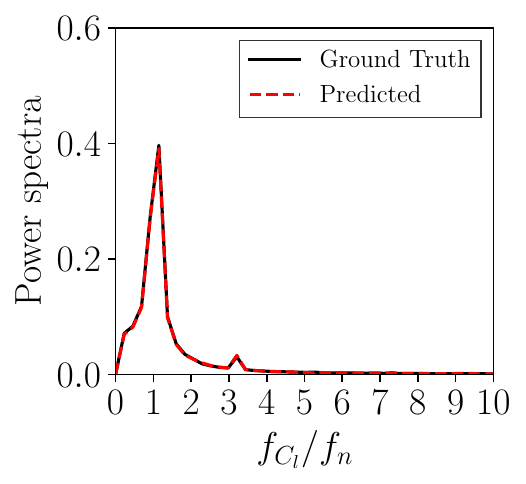}\\
    \caption{}
\end{subfigure}
\caption{\label{cl-cd-VIV}Comparison of the hydrodynamic coefficient predictions with ground truth data in isolated configuration at a test \(U^* = 5.125\). Subplots (a) and (c) show the time histories of the fluctuating drag and lift coefficients, respectively. Subplots (b) and (d) present the corresponding power spectra.}
\end{figure*}

Following the training of the mesh prediction sub-network, we focus on assessing the hypergraph neural network's accuracy in capturing the flow field evolution during the cylinder's oscillatory dynamics. 
A detailed comparison between the predicted and ground truth flow fields of normalized pressure ($2p / \rho^{f} U_0^2$), streamwise velocity ($u_x / U_0$), and crossflow velocity ($u_y / U_0$) are presented in Fig.~\ref{contours-VIV} at a test reduced velocity of $U^* = 5.125$. The predicted fields qualitatively replicate the major flow features and wake dynamics observed in the ground truth. The prediction errors are generally localized in regions near the wake centerline and remain low across all fields.
To quantitatively assess the model’s performance on key hydrodynamic metrics, we have computed the fluctuating drag and lift coefficients, denoted by \(C_d - \overline{C_d}\) and \(C_l - \overline{C_l}\) respectively, using the predicted and ground truth flow fields. Figure~\ref{cl-cd-VIV} presents a comparison of the coefficients between the predictions and full-order simulation results at $U^* = 5.125$. The predictions agree well with the ground truth in both the temporal and spectral domains. The model is found to accurately predict the dominant vortex shedding frequencies, as seen by the alignment of the peak spectral content in Fig.~\ref{cl-cd-VIV}. Minor discrepancies are observed in the drag coefficient predictions; however, the dominant features of the force signals are accurately predicted.

\begin{figure*}
\centering
\begin{subfigure}{0.48\textwidth}\centering
    \includegraphics[width=1\linewidth]{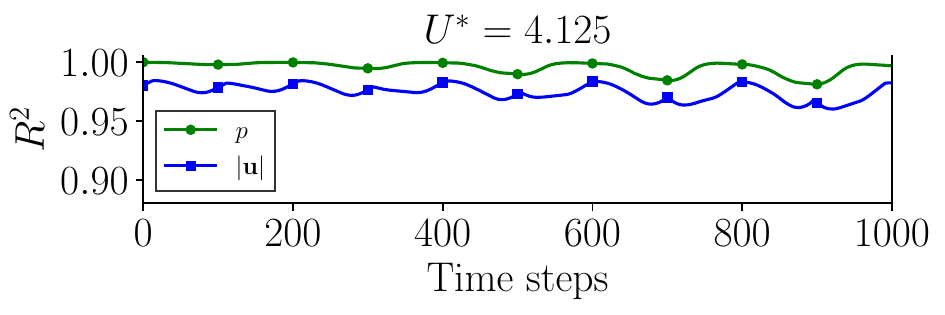}
    \caption{}
\end{subfigure}
\begin{subfigure}{0.48\textwidth}\centering
    \includegraphics[width=1\linewidth]{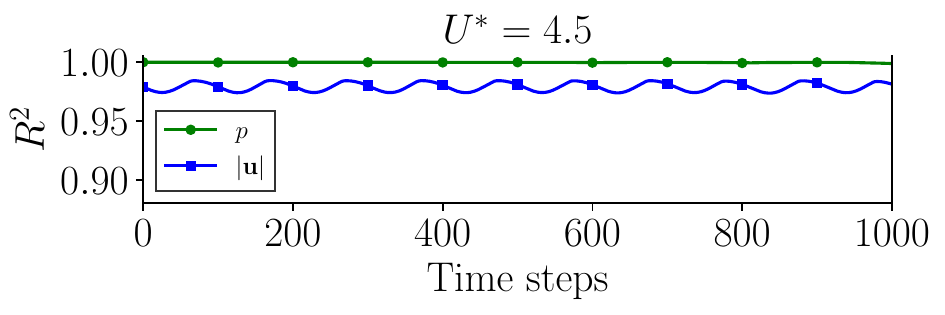}
    \caption{}
\end{subfigure}\\
\begin{subfigure}{0.48\textwidth}\centering
    \includegraphics[width=1\linewidth]{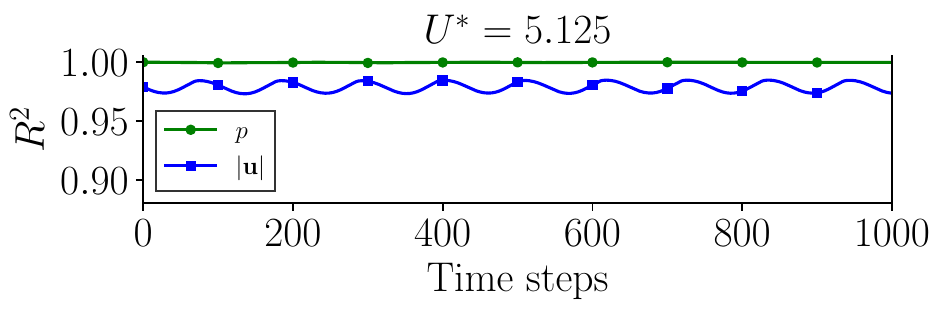}
    \caption{}
\end{subfigure}
\begin{subfigure}{0.48\textwidth}\centering
    \includegraphics[width=1\linewidth]{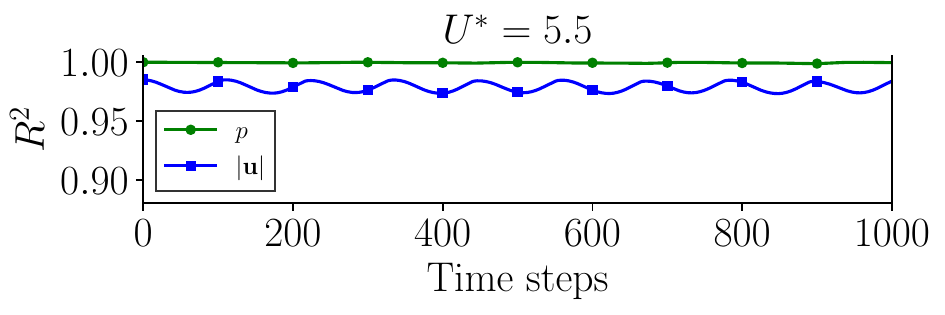}
    \caption{}
\end{subfigure}\\
\begin{subfigure}{0.48\textwidth}\centering
    \includegraphics[width=1\linewidth]{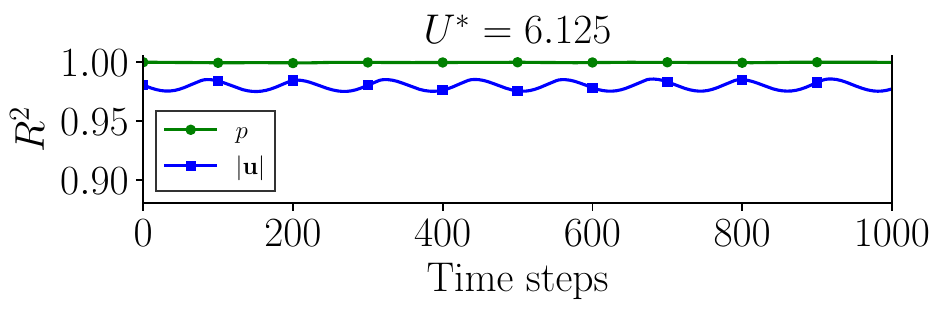}
    \caption{}
\end{subfigure}
\begin{subfigure}{0.48\textwidth}\centering
    \includegraphics[width=1\linewidth]{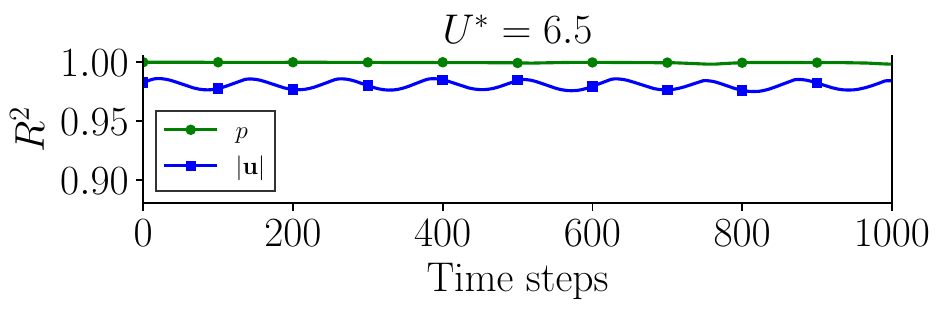}
    \caption{}
\end{subfigure}
\caption{\label{r2-VIV}Coefficient of determination
between ground truth and predicted flow fields at every prediction time step in the isolated configuration.}
\end{figure*}

\begin{figure*}
\centering
\begin{subfigure}{0.49\textwidth}\centering
    \includegraphics[width=1\linewidth]{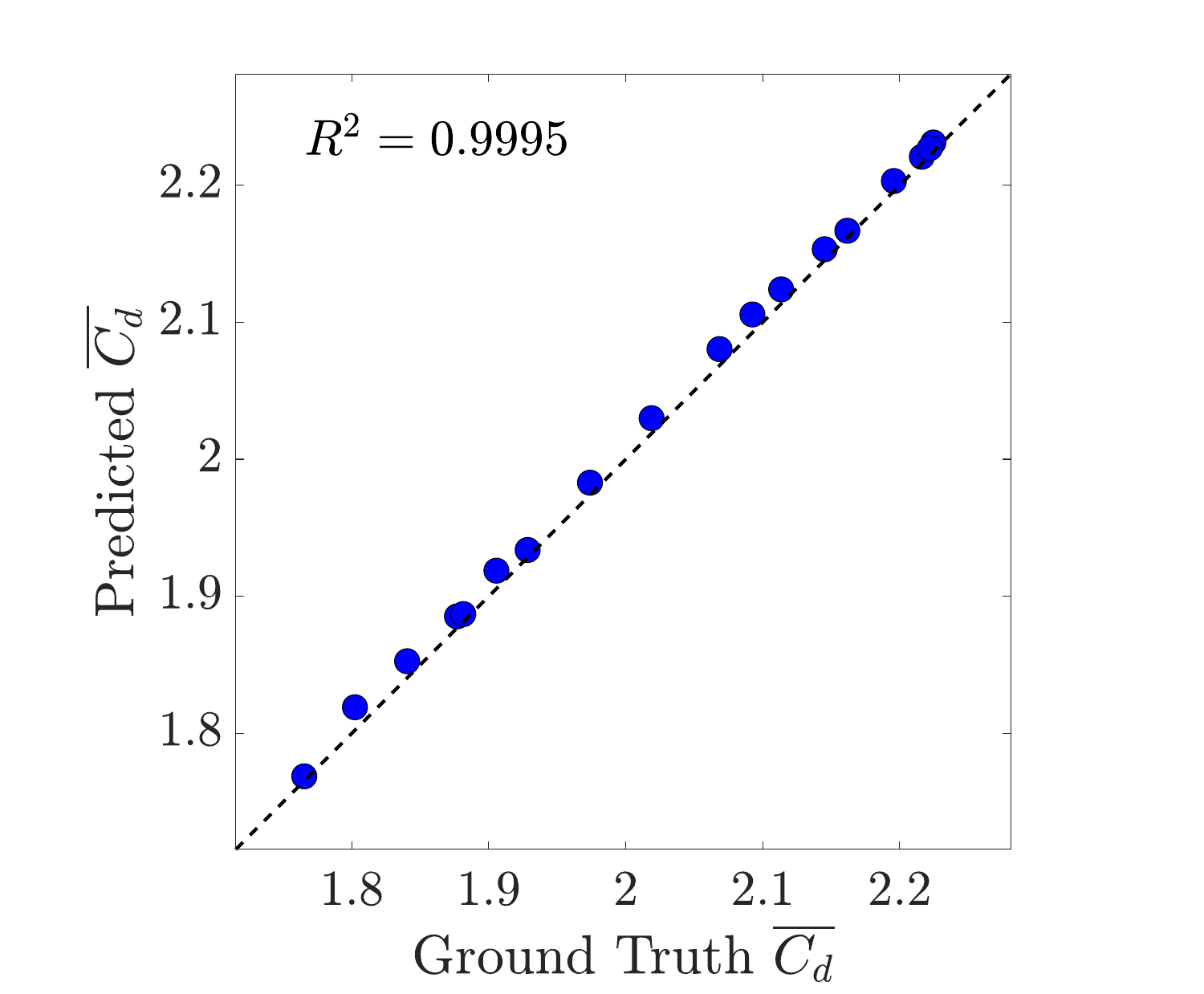}
    \caption{}
\end{subfigure}
\begin{subfigure}{0.49\textwidth}\centering
    \includegraphics[width=1\linewidth]{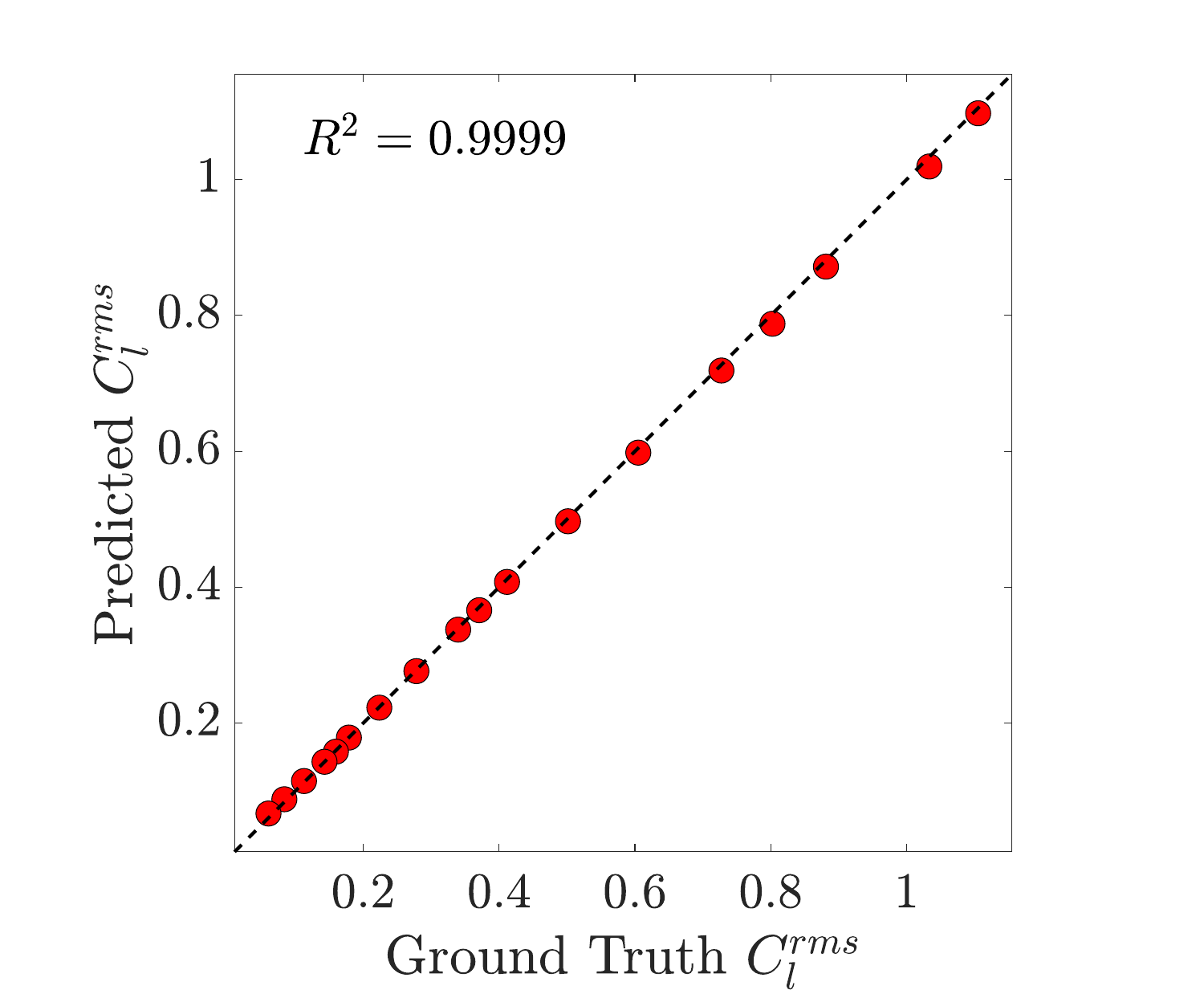}
    \caption{}
\end{subfigure}
\caption{\label{Cd-Cl-Re60-VIV}Comparison between the ground truth and predicted values of (a) the mean drag coefficient \( \overline{C_d} \) and (b) root mean square fluctuating lift coefficient \( C_l^{{rms}} \) in isolated configuration.}
\end{figure*}

The temporal evolution of the coefficient of determination $R^2$ for the pressure and velocity magnitude across multiple reduced velocities is shown in Fig.~\ref{r2-VIV} during the prediction rollout. The coefficient of determination is specified as:
\begin{align}
R^2 = 1 - \frac{\| \mathbf{q} - \hat{\mathbf{q}} \|_2^2}{\| \mathbf{q} - \bar{\mathbf{q}} \|_2^2},
\end{align}
where $\mathbf{q}$ denotes the true system state, $\hat{\mathbf{q}}$ the predicted state, and $\bar{\mathbf{q}}$ the mean of the true state. In all test cases, the model maintains high predictive accuracy, with $R^2$ values consistently above 0.95. 
Further quantitative evaluation of the flow field prediction sub-network is done by comparing the predicted values of the mean drag coefficient \(\overline{{C}_d}\) and the root mean square fluctuating lift coefficient \(C_l^{{rms}}\) against the ground truth results across the whole dataset. The results are given in Fig.~\ref{Cd-Cl-Re60-VIV}.
The predicted values are shown to agree well with the reference values, with coefficients of determination exceeding \(R^2 \geq 0.9995\). 

In the following, we extend the analysis to a related case involving an oscillating cylinder in a tandem configuration. This setup features intricate wake–vortex interactions that lead to multi-frequency response dynamics in the cylinder. The focus is on evaluating the prediction capabilities of the framework under wake interference effects.


\subsection{Prediction of FIV Dynamics in Tandem Cylinders}
In the tandem configuration cases, a rigid stationary cylinder with identical geometry is positioned upstream at a streamwise distance \(x_0\) from the freely oscillating cylinder. The separation distance between the cylinders is set to \(x_0 = 5D\), which exceeds the critical threshold for bistable shear layer reattachment. This setup enables the development of vortex shedding in the gap region between the two cylinders. We begin by assessing the capability of the trained mesh prediction sub-network in capturing the oscillatory response of the downstream cylinder.

\begin{figure*}
\centering
\begin{subfigure}{0.5\textwidth}\centering
    \includegraphics[width=1\linewidth]{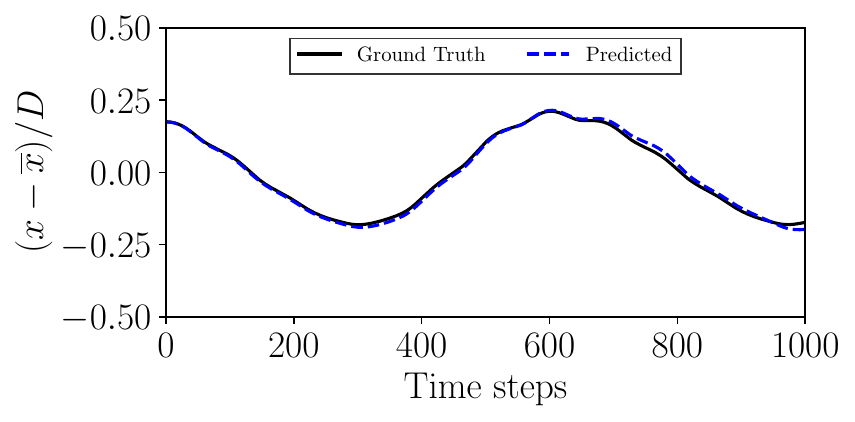}
    \caption{}
\end{subfigure}
\begin{subfigure}{0.27\textwidth}\centering
    \includegraphics[width=1\linewidth]{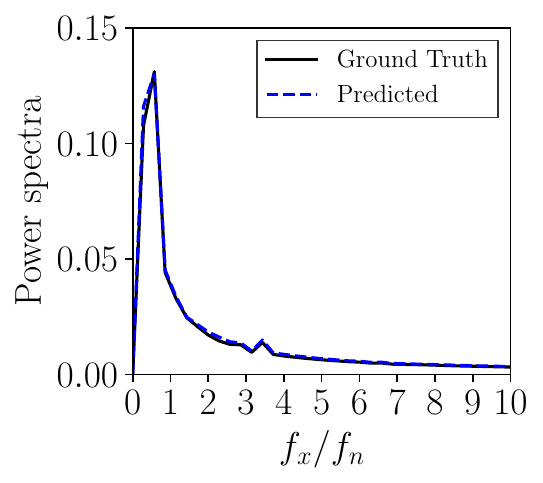}\\
    \caption{}
\end{subfigure}
\begin{subfigure}{0.5\textwidth}\centering
    \includegraphics[width=1\linewidth]{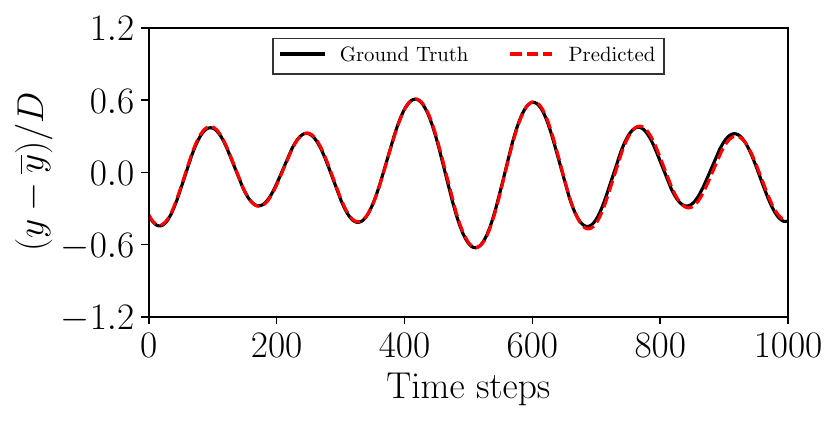}
    \caption{}
\end{subfigure}
\begin{subfigure}{0.28\textwidth}\centering
    \includegraphics[width=1\linewidth]{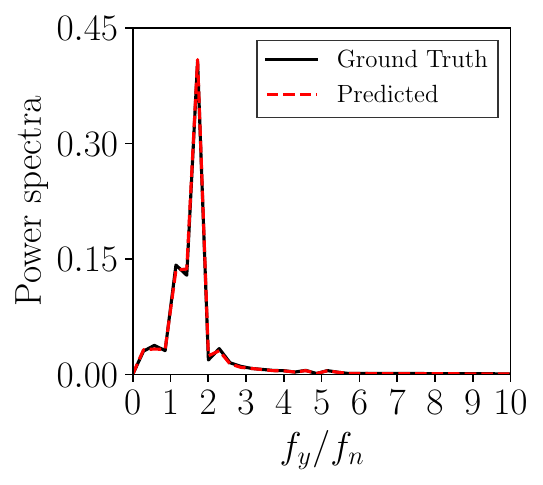}\\
    \caption{}
\end{subfigure}
\caption{\label{time_histories_tandem}Comparison of the downstream cylinder displacement predictions with ground truth data in tandem configuration at a test $Re=70$. Subplots (a)
and (c) show the time histories of the normalized streamwise and transverse displacements, respectively. Subplots (b) and (d) present the corresponding power spectra.}
\end{figure*}

\begin{figure*}
\centering
\begin{subfigure}{0.32\textwidth}\centering
    \includegraphics[width=0.98\linewidth]{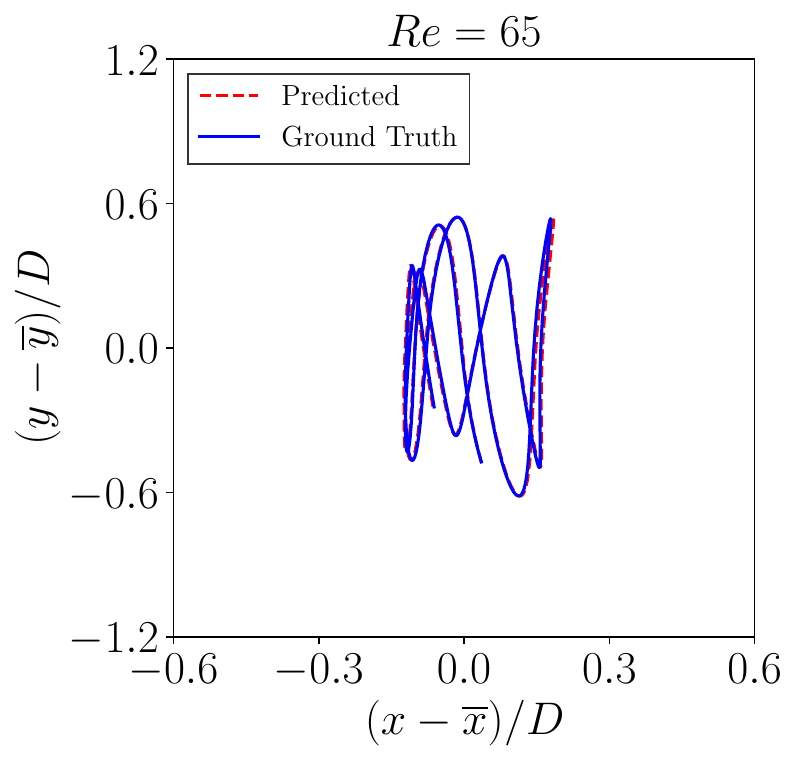}
    \caption{}
\end{subfigure}
\begin{subfigure}{0.32\textwidth}\centering
    \includegraphics[width=0.98\linewidth]{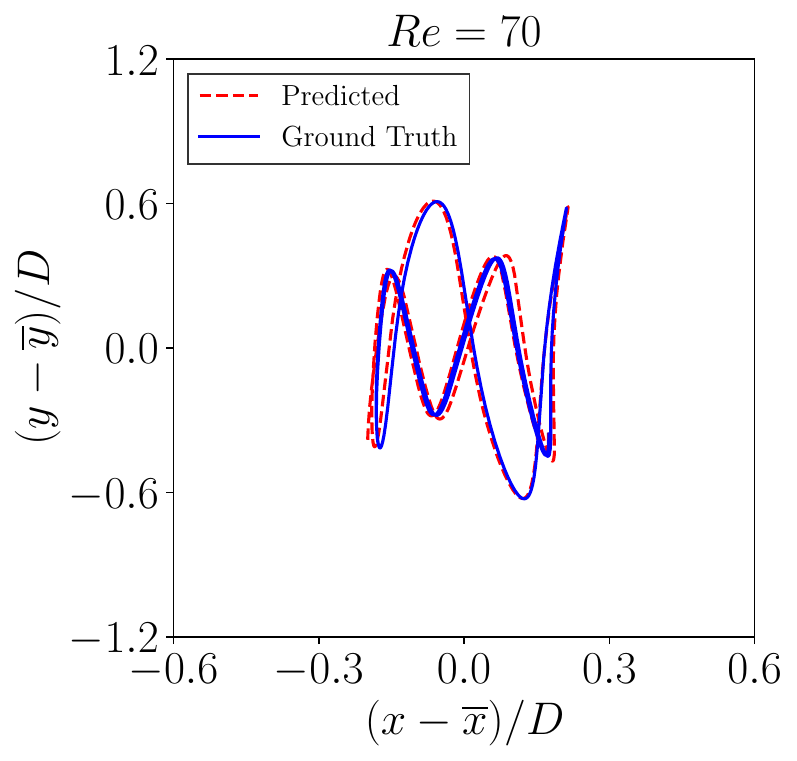}
    \caption{}
\end{subfigure}\\
\begin{subfigure}{0.32\textwidth}\centering
    \includegraphics[width=0.98\linewidth]{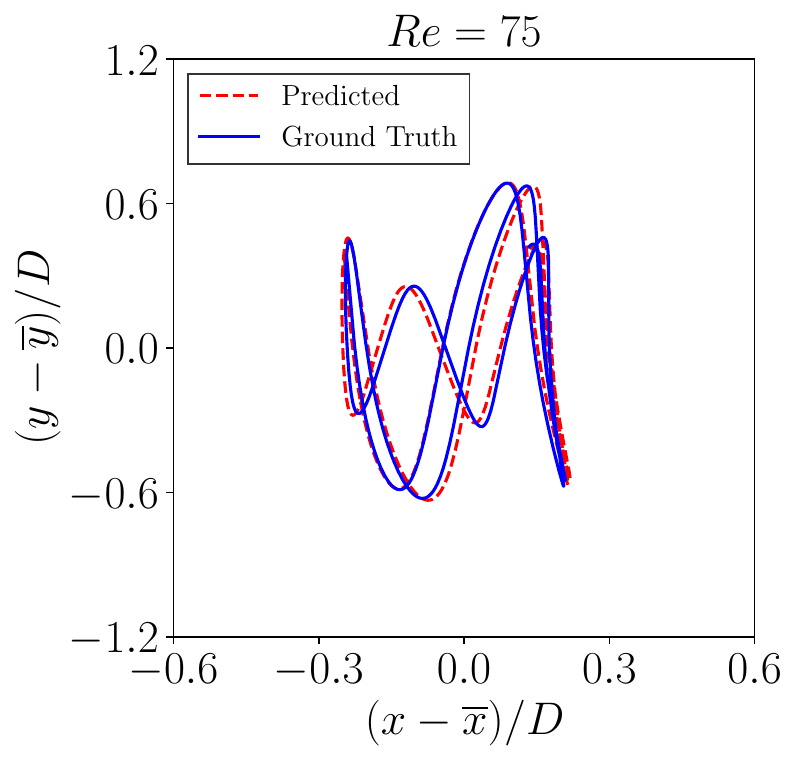}
    \caption{}
\end{subfigure}
\begin{subfigure}{0.32\textwidth}\centering
    \includegraphics[width=0.98\linewidth]{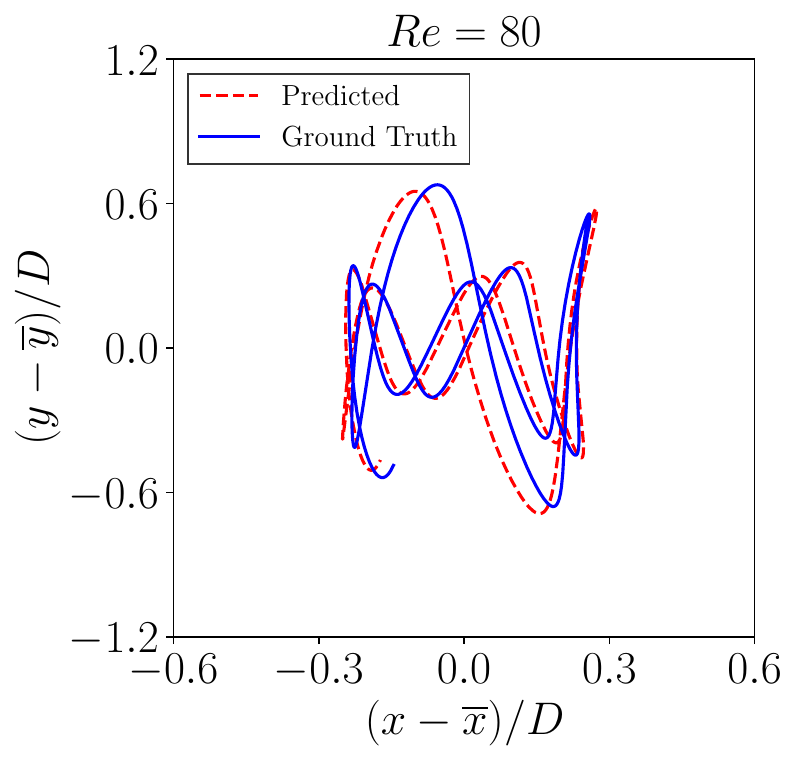}
    \caption{}
\end{subfigure}
\caption{\label{trajectories-tandem}Comparison of downstream cylinder's motion trajectories between predictions and ground truth results for selected cases in tandem configuration.}
\end{figure*}

Figure~\ref{time_histories_tandem} presents a comparison of the time histories and corresponding power spectra of the cylinder's streamwise and transverse displacements at a test \(Re = 70\). As it is seen, the predictions closely match the ground-truth data over the roll-out time steps. The trained model demonstrates accurate predictions of frequency peaks and reconstruction of the dominant oscillatory response.
The motion trajectories of the oscillating downstream cylinder over roll-out predictions in the \(x\)-\(y\) plane are presented in Fig.~\ref{trajectories-tandem} for select cases. These cases lie within the upper end of the Reynolds number range characterized by increased vortex shedding complexity and modulation in the structural response. Unlike the figure-eight patterns observed in the isolated configuration, the trajectories here exhibit intricate, multi-looped orbits reflecting stronger nonlinear coupling and wake interference effects. The model effectively captures the global topology and phase relationships of the oscillatory motion. Although small local deviations are observed, particularly in loop overlap, the predicted paths retain the key dynamic features of the true system response.

\begin{figure*}
\centering
\begin{subfigure}{0.328\textwidth}\centering
    \includegraphics[width=1\linewidth]{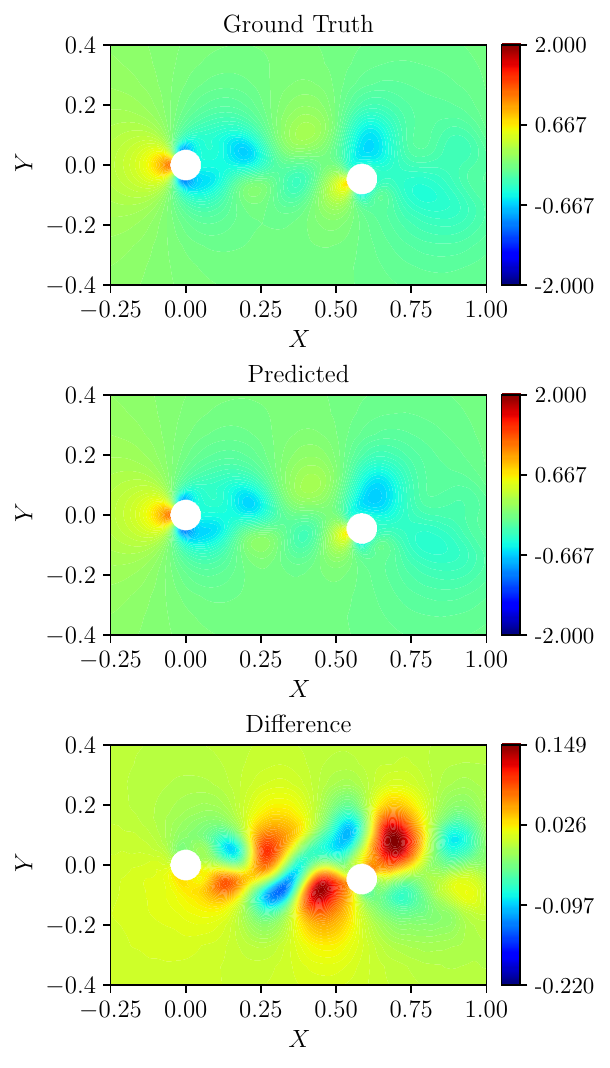}
    \caption{}
\end{subfigure}
\begin{subfigure}{0.328\textwidth}\centering
    \includegraphics[width=1\linewidth]{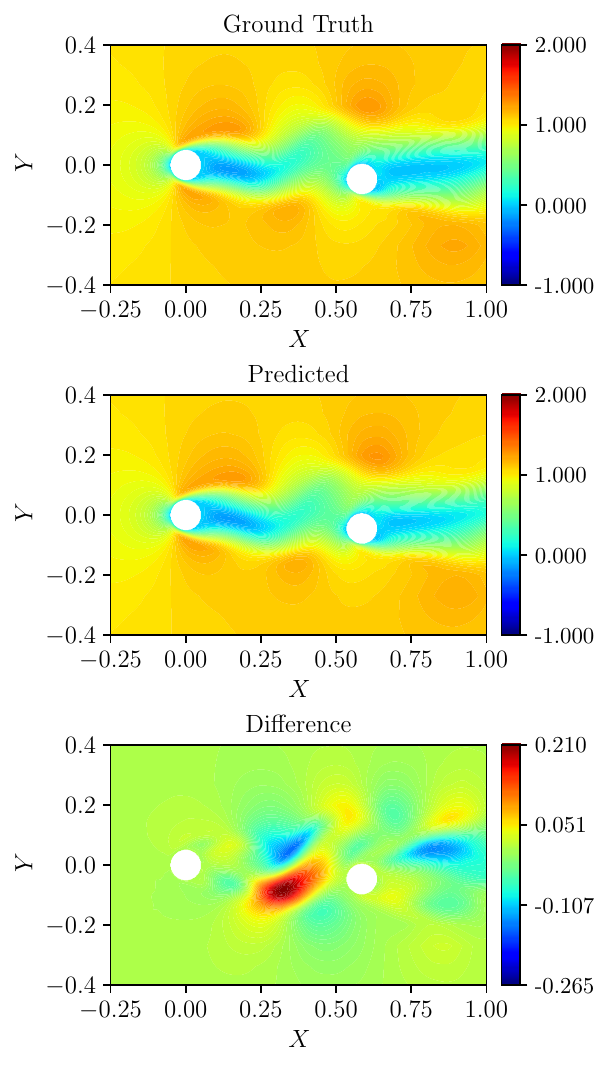}
    \caption{}
\end{subfigure}
\begin{subfigure}{0.328\textwidth}\centering
    \includegraphics[width=1\linewidth]{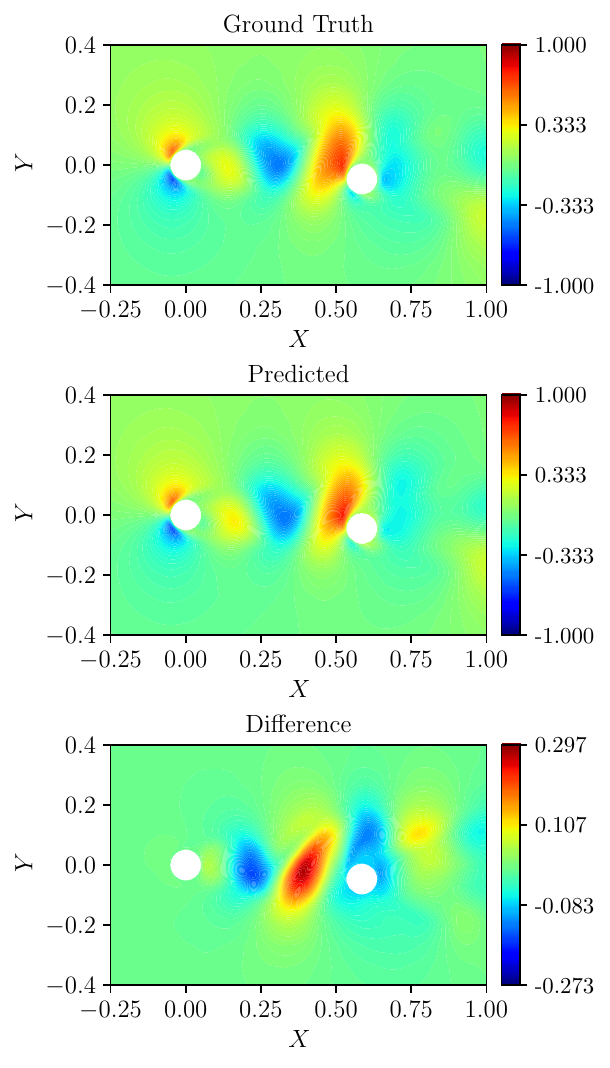}\\
    \caption{}
\end{subfigure}
\caption{\label{contours-Tandem}Comparison between predicted and ground truth contours of normalized (a) pressure ($2p/\rho^{f}U_{0}^2$), (b) streamwise velocity ($u_{x}/U_{0}$), and (c) crossflow velocity ($u_{y}/U_{0}$) for the tandem configuration at a test \(Re = 70\); fields are shown at roll‑out step 1000.}
\end{figure*}

\begin{figure*}
\centering
\begin{subfigure}{0.49\textwidth}\centering
    \includegraphics[width=1\linewidth]{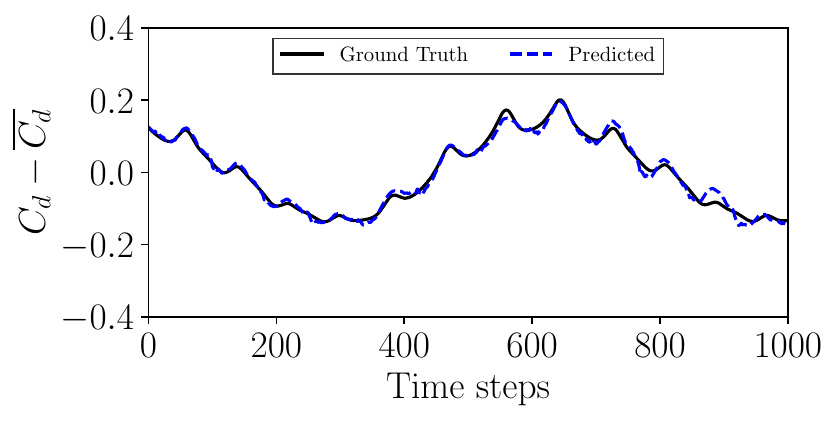}
    \caption{}
\end{subfigure}
\begin{subfigure}{0.28\textwidth}\centering
    \includegraphics[width=1\linewidth]{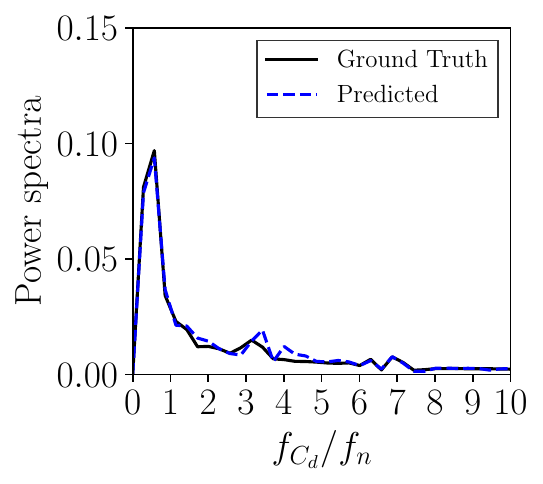}\\
    \caption{}
\end{subfigure}
\begin{subfigure}{0.49\textwidth}\centering
    \includegraphics[width=1\linewidth]{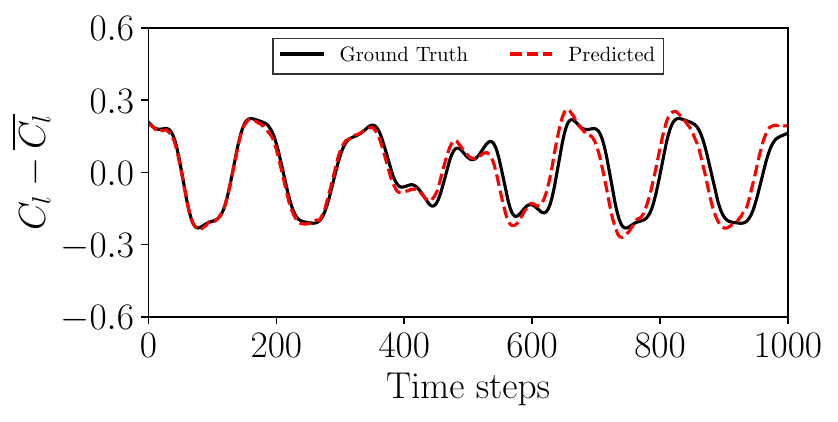}
    \caption{}
\end{subfigure}
\begin{subfigure}{0.27\textwidth}\centering
    \includegraphics[width=1\linewidth]{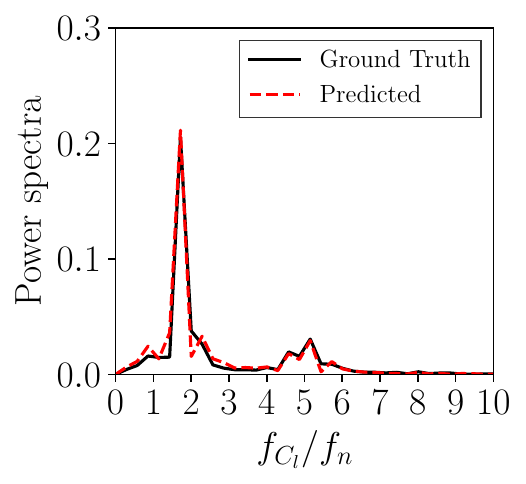}\\
    \caption{}
\end{subfigure}
\caption{\label{time_histories_force_tandem}Comparison of the hydrodynamic coefficient predictions with ground truth data in tandem configuration at a test $Re=70$. Subplots (a)
and (c) show the time histories of the fluctuating drag and lift coefficients, respectively. Subplots
(b) and (d) present the corresponding power spectra.  The coefficients are computed on the surface of the oscillating cylinder.}
\end{figure*}

With the trained mesh prediction sub-network established, we proceed to evaluate the flow field prediction sub-network. Figure~\ref{contours-Tandem} presents a comparison between the predicted and ground truth contours of normalized pressure and velocity for the test \(Re = 70\). The spatial structure and magnitude of the predicted fields show strong agreement with the reference data. Discrepancies are primarily localized in the near-wake region between the two cylinders, where strong unsteady wake interference creates highly nonlinear flow dynamics. Despite these localized deviations, the predicted fields preserve the key dynamic characteristics of the flow.
The time series and power spectra of the fluctuating drag and lift coefficients are extracted from the predicted and full‑order flow fields at $Re=70$ and presented in Fig.~\ref{time_histories_force_tandem}. The fluctuating drag coefficient, characterized by a more irregular and broadband signature, is reproduced with strong temporal alignment despite moderate amplitude modulation. The fluctuating lift coefficient is also predicted with good accuracy. 

Closer inspection of the lift coefficient variations reveals minor discrepancies toward the later part of the roll-out, where slight phase shifts and amplitude variations are observed near the peaks. Mitigation of the observed deviations is possible by anchoring the prediction to true system states at intermediate checkpoints during inference. This strategy is particularly relevant for applications involving sensor data fusion at fixed intervals, where intermittent access to ground truth measurements can help correct cumulative prediction drift. 

\begin{figure*}
\centering
\begin{subfigure}{0.48\textwidth}\centering
    \includegraphics[width=1\linewidth]{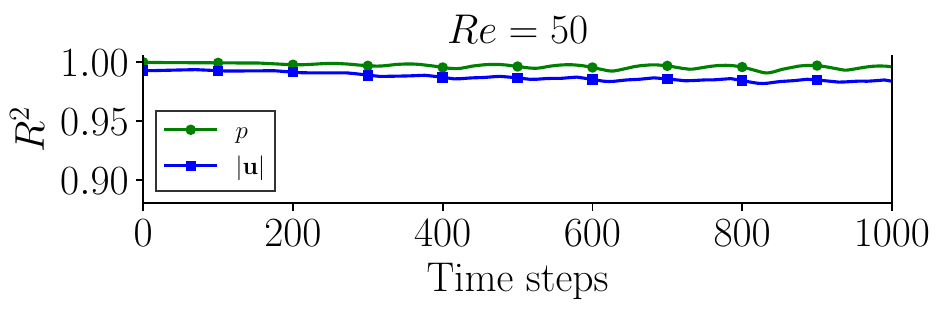}
    \caption{}
\end{subfigure}
\begin{subfigure}{0.48\textwidth}\centering
    \includegraphics[width=1\linewidth]{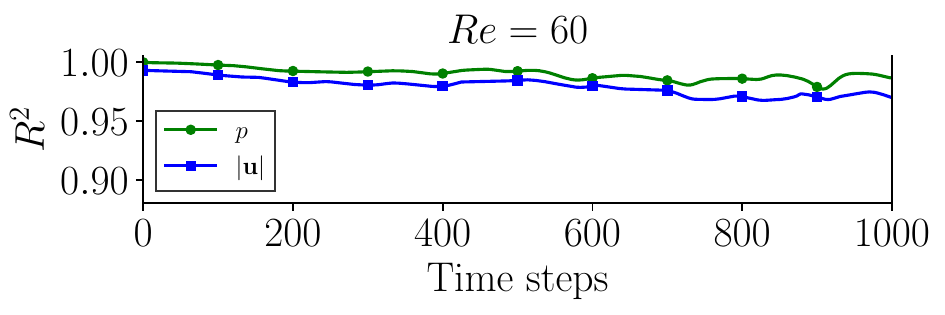}
    \caption{}
\end{subfigure}
\begin{subfigure}{0.48\textwidth}\centering
    \includegraphics[width=1\linewidth]{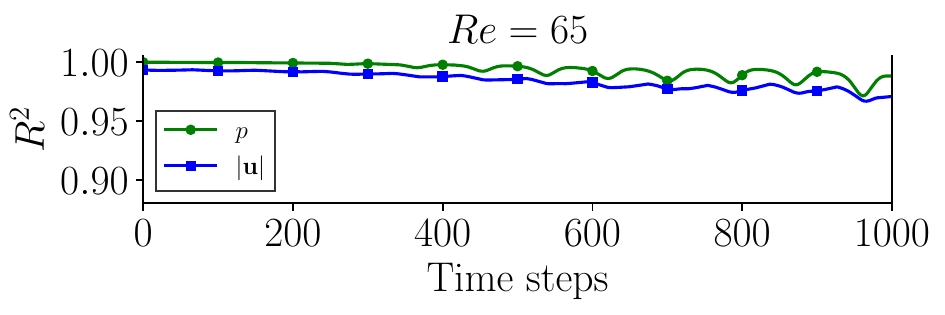}
    \caption{}
\end{subfigure}
\begin{subfigure}{0.48\textwidth}\centering
    \includegraphics[width=1\linewidth]{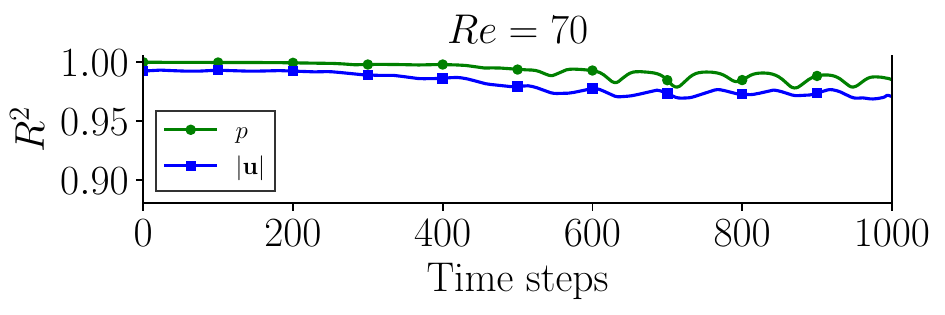}
    \caption{}
\end{subfigure}\\
\begin{subfigure}{0.48\textwidth}\centering
    \includegraphics[width=1\linewidth]{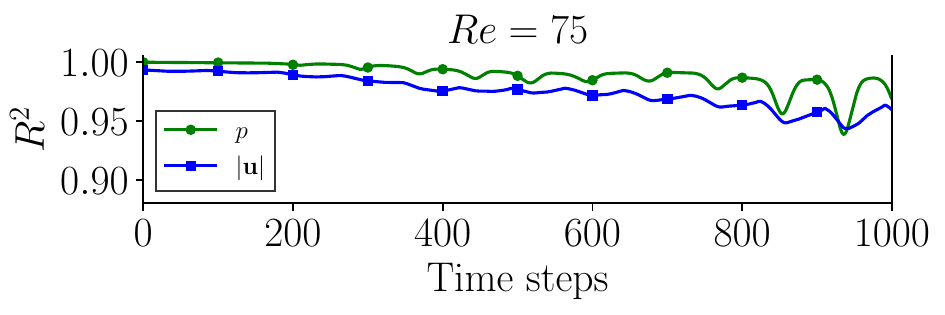}
    \caption{}
\end{subfigure}
\begin{subfigure}{0.48\textwidth}\centering
    \includegraphics[width=1\linewidth]{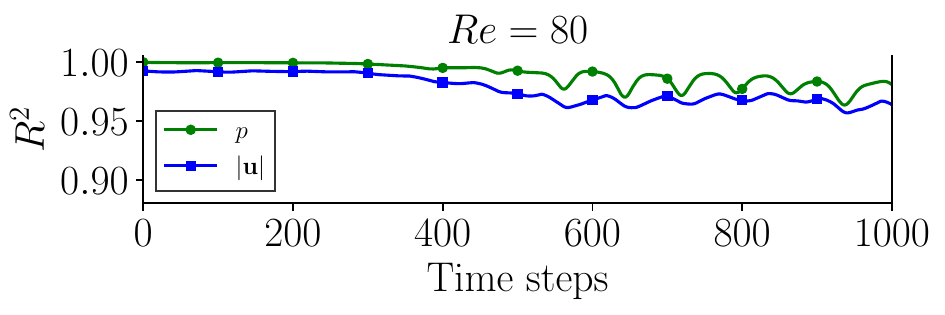}
    \caption{}
\end{subfigure}
\caption{\label{r2}Coefficient of determination
between ground truth and predicted flow fields at every prediction time step in the tandem configuration.}
\end{figure*}

Figure~\ref{r2} shows the coefficient of determination $R^2$ values for pressure and velocity magnitude over time across a range of Reynolds numbers in the tandem configuration. The model sustains high $R^2$ values throughout the rollout, reflecting strong predictive performance. A gradual decline in $R^2$ is observed as time progresses, which reflects the increasing challenge of resolving complex, unsteady wake interactions over long prediction horizons. Despite this, the overall consistency in $R^2$ highlights the model’s ability to preserve physical coherence across examined cases. To mitigate the temporal decay in prediction accuracy, one could consider anchoring the rollout to ground truth data at intermediate intervals, thereby reinforcing stability and long-term fidelity.

\begin{figure*}
\centering
\begin{subfigure}{0.49\textwidth}\centering
    \includegraphics[width=1\linewidth]{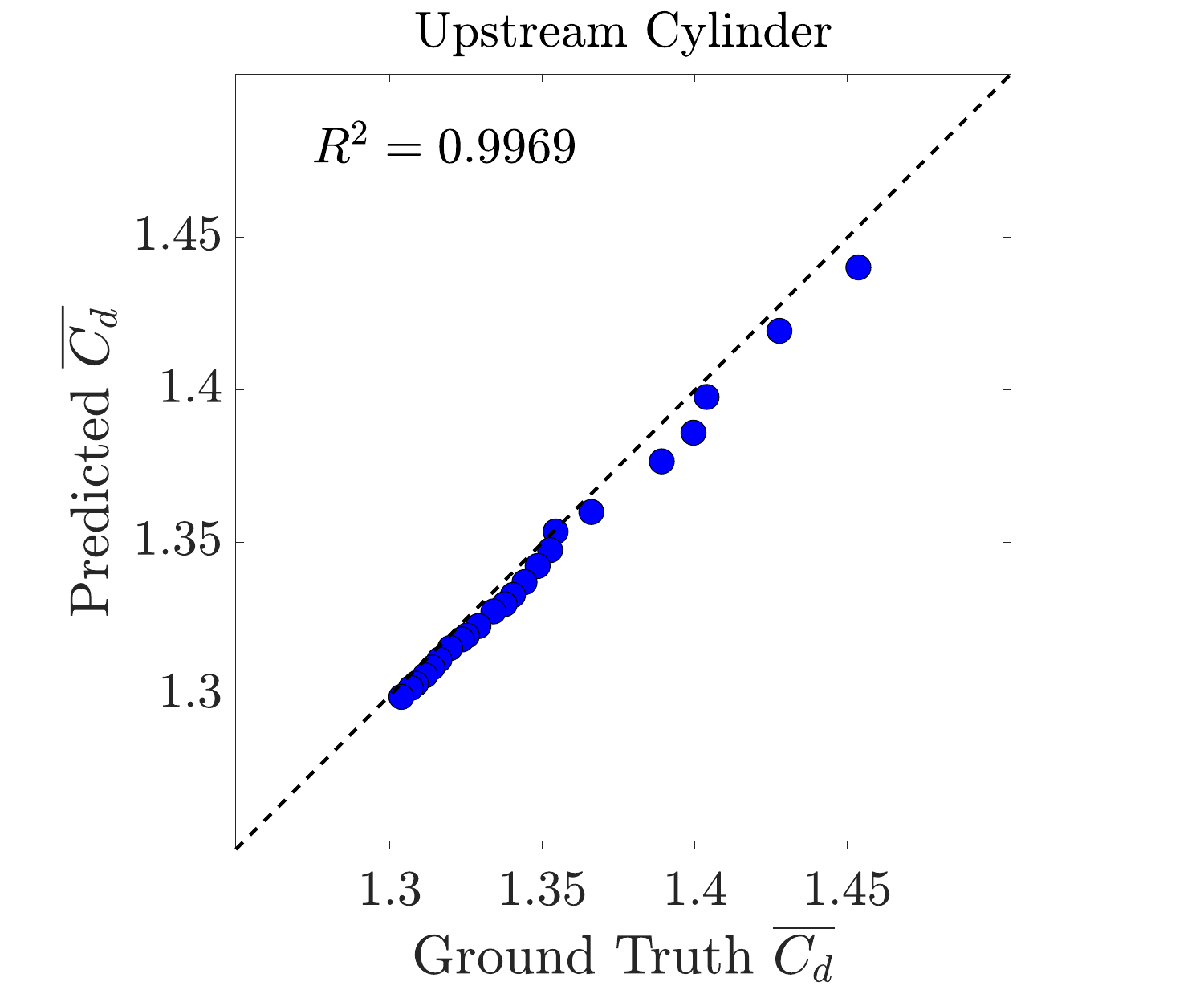}
    \caption{}
\end{subfigure}
\begin{subfigure}{0.49\textwidth}\centering
    \includegraphics[width=1\linewidth]{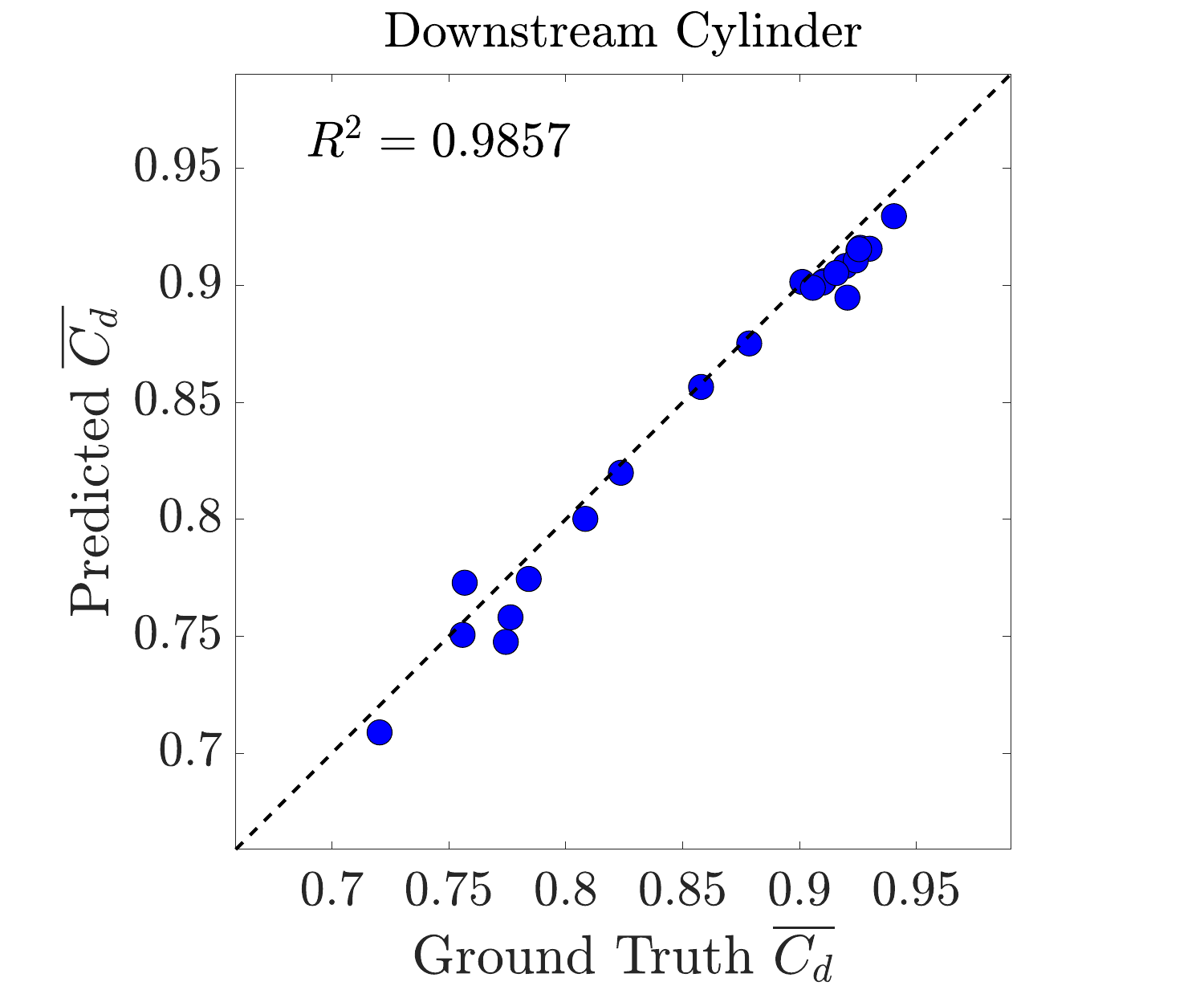}
    \caption{}
\end{subfigure}\\
\begin{subfigure}{0.49\textwidth}\centering
    \includegraphics[width=1\linewidth]{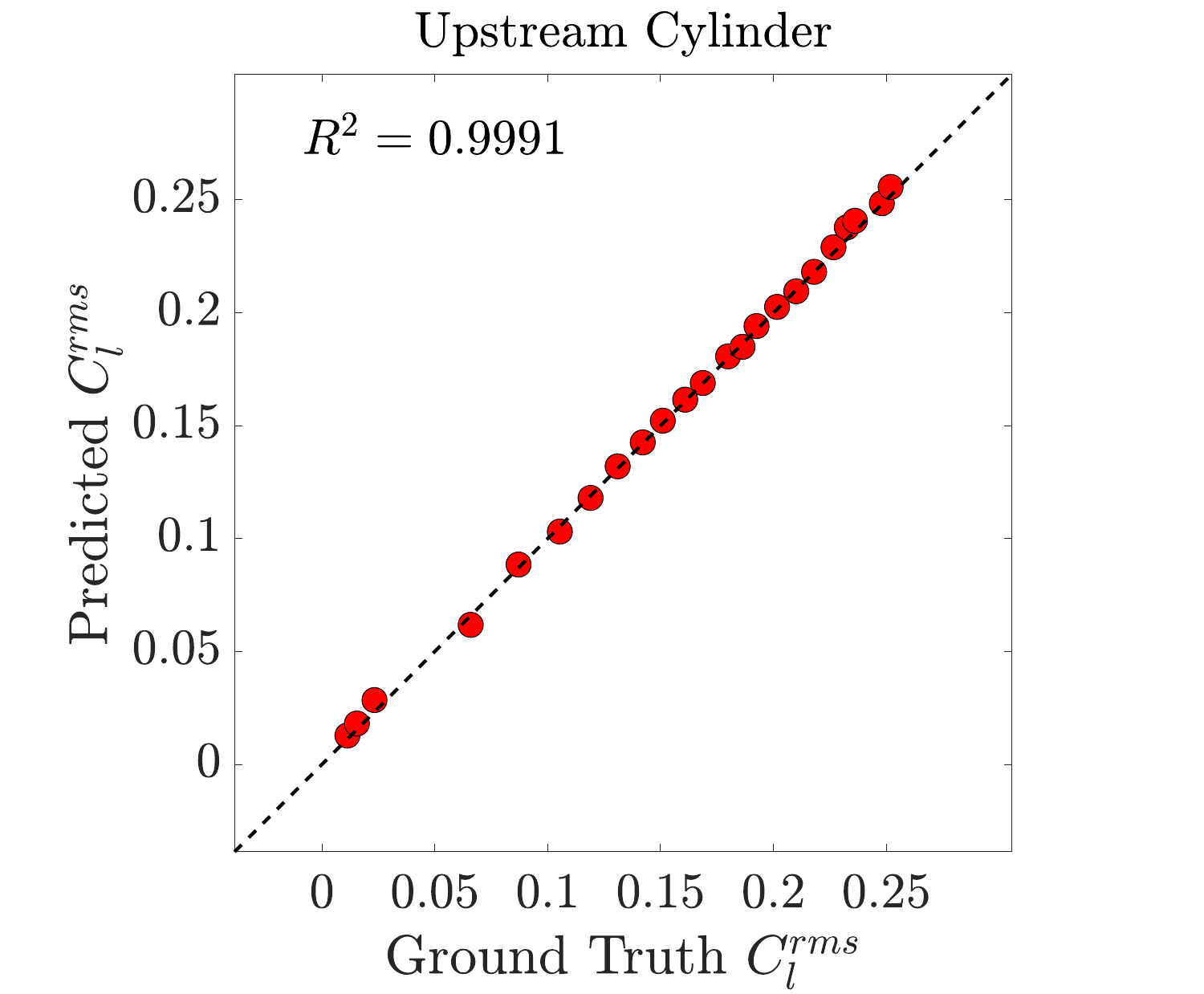}
    \caption{}
\end{subfigure}
\begin{subfigure}{0.49\textwidth}\centering
    \includegraphics[width=1\linewidth]{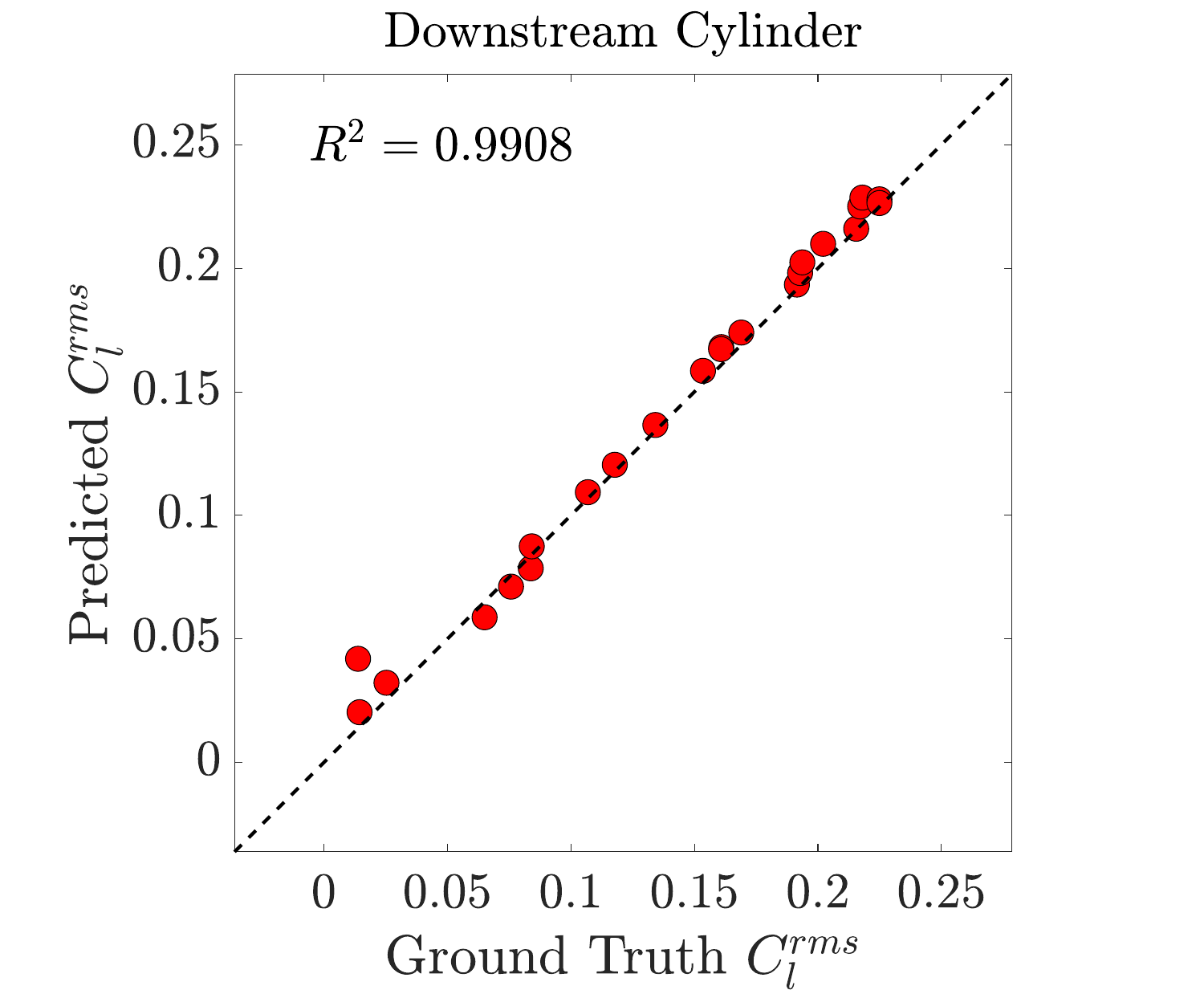}
    \caption{}
\end{subfigure}
\caption{\label{Ay-Cl-Ustar9-Tandem}Comparison between the ground truth and predicted values of (a, b) the mean drag coefficient $\overline{C_d}$, and (c, d) the root mean square fluctuating lift coefficient $C_l^{{rms}}$, for the upstream and downstream cylinders in tandem configuration. The results include the training and test datasets.}
\end{figure*}

Figure~\ref{Ay-Cl-Ustar9-Tandem} presents a comparison between predicted and ground truth values of the mean drag coefficient $\overline{C_d}$ and the root mean square fluctuating lift coefficient $C_l^{{rms}}$ for the upstream and downstream cylinders in the tandem configuration. The model demonstrates excellent agreement with the reference data, achieving $R^2>0.99$ for $\overline{C_d}$ and $C_l^{{rms}}$ on the upstream cylinder, and slightly lower, but still strong, agreement on the downstream cylinder with $R^2>0.98$. These results confirm that the model accurately captures force statistics across both bodies, even in the presence of complex wake-body interactions characteristic of tandem configurations.

\section{Conclusion}
\label{sec:conclusion}
In this work, we presented a hypergraph neural network framework for predicting flow-induced vibrations in freely oscillating cylinders. The framework constructs node-element hypergraphs from unstructured computational meshes, enabling the encoding of higher-order relational information inherent to the dynamics of fluid–structure interactions. We adopted a modular, partitioned architecture comprising two specialized sub-networks: a complex-valued proper orthogonal decomposition network that captures mesh deformation through a low-rank representation of Arbitrary Lagrangian-Eulerian grid motion, and a hypergraph-based message-passing network that predicts the evolution of unsteady fluid flow fields. The framework was validated using high-fidelity simulation data across isolated and tandem cylinder configurations, spanning a range of Reynolds numbers and reduced velocities. 
The framework demonstrated strong predictive capabilities and accurately reconstructed hydrodynamic force coefficients, oscillation amplitudes, and dominant frequencies, confirming its ability to preserve fidelity across spatial and temporal scales. 
While our primary aim was to explore the efficacy of finite element-inspired hypergraph neural networks, systematic assessments with other architectures of graph neural networks can be explored.
Future work could also focus on enhancing the interpretability and decision-making capacity of the framework by incorporating explainable graph neural network architectures and extending evaluations to high Reynolds number flows. These advancements will pave the way for the development of digital twins and support reliable deployment in complex engineering systems such as those found in marine, offshore, and aerospace domains. Overall, this study marks a step toward scalable, data-driven modeling of multiphysics systems with potential for real-time monitoring and control applications.

\section*{Acknowledgment}
The authors would like to acknowledge the Natural Sciences and Engineering Research Council of Canada (NSERC) for funding the project. The research was enabled in part through computational resources and services provided by the Digital Research Alliance of Canada and the Advanced Research Computing facility at the University of British Columbia.


\bibliography{references}

\end{document}